\documentclass[aps,twocolumn,groupedaddress]{revtex4}
\usepackage{graphicx}  
\usepackage{dcolumn}   
\usepackage{bm}        
\usepackage{amssymb}   
\usepackage{multirow}
\usepackage{endnotes}
\usepackage{amsmath, amsthm}
\usepackage[all]{xy}
\usepackage{verbatim}
\usepackage{color}
\usepackage{array,hhline,dcolumn} 
\usepackage[nice]{nicefrac}

\def\th{\theta}

\def\si{\sigma}

\def\De{\Delta}

\def\anu{\alpha_\nu}
\def\anub{\alpha_{\bar{\nu}}}
\def\nubar{\bar{\nu}}
\def\tmu{T_\mu}
\def\nub{\bar{\nu}}

\def\to{\rightarrow}

\def\beq{\begin{eqnarray}}
\def\eeq{\end{eqnarray}}

\def\ucmt{\mbox{cm}^2}

\def\uGeV{\mbox{GeV}}

\def\nuance{\textsc{nuance}}

\def\nue{\nu_e}
\def\nueb{\bar{\nu}_e}
\def\numu{\nu_\mu}
\def\numub{\bar{\nu}_\mu}

\def\elp{e^{+}}

\def\mup{\mu^{+}}
\def\mum{\mu^{-}}
\def\piz{\pi^{0}}

\def\pip{\pi^{+}}
\def\pim{\pi^{-}}
\def\thetmu{\theta_{\mu}}

\def\MBosc1POT{5.58\times 10^{20}}

\def\uz{\,\textrm{cos}\, \theta_\mu}
\def\qsq{Q^2}
\def\qsqqe{Q^2_{QE}}
\def\enuqe{E_{\nu}^{QE}}
\def\dbldiffl{\frac{d^2\sigma}{dT_\mu d\uz}}
\def\pipm{\pi^{\pm}}
\def\maeffc{M_{A}^{\mathrm{eff,C}}}
\def\maeffh{M_{A}^{\mathrm{eff,H}}}

\begin{document}


\title{First Measurement of the Muon Antineutrino Double-Differential Charged Current Quasi-Elastic Cross Section}
\date{\today}
\author{
        A.~A. Aguilar-Arevalo$^{12}$, 
        B.~C.~Brown$^{6}$, L.~Bugel$^{11}$,
	G.~Cheng$^{5}$, E.~D.~Church$^{15}$, J.~M.~Conrad$^{11}$,
	R.~Dharmapalan$^{1}$, 
	Z.~Djurcic$^{2}$, D.~A.~Finley$^{6}$, R.~Ford$^{6}$,
        F.~G.~Garcia$^{6}$, G.~T.~Garvey$^{9}$, 
        J.~Grange$^{7}$,
        W.~Huelsnitz$^{9}$, C.~Ignarra$^{11}$, R.~Imlay$^{10}$,
        R.~A. ~Johnson$^{3}$, G.~Karagiorgi$^{5}$, T.~Katori$^{11}$,
        T.~Kobilarcik$^{6}$, 
        W.~C.~Louis$^{9}$, C.~Mariani$^{16}$, W.~Marsh$^{6}$,
        G.~B.~Mills$^{9}$,
	J.~Mirabal$^{9}$,
        C.~D.~Moore$^{6}$, J.~Mousseau$^{7}$, 
        P.~Nienaber$^{14}$, 
        B.~Osmanov$^{7}$, Z.~Pavlovic$^{9}$, D.~Perevalov$^{6}$,
        C.~C.~Polly$^{6}$, H.~Ray$^{7}$, B.~P.~Roe$^{13}$,
        A.~D.~Russell$^{6}$, 
	M.~H.~Shaevitz$^{5}$, 
        J.~Spitz$^{11}$, I.~Stancu$^{1}$, 
        R.~Tayloe$^{8}$, R.~G.~Van~de~Water$^{9}$, M.~O.~Wascko$^{}$,
        D.~H.~White$^{9}$, D.~A.~Wickremasinghe$^{3}$, G.~P.~Zeller$^{6}$,
        E.~D.~Zimmerman$^{4}$ \\
\smallskip
(MiniBooNE Collaboration)
\smallskip
}
\smallskip
\smallskip
\affiliation{
$^1$University of Alabama; Tuscaloosa, AL 35487 \\
$^2$Argonne National Laboratory; Argonne, IL 60439 \\
$^3$University of Cincinnati; Cincinnati, OH 45221\\
$^4$University of Colorado; Boulder, CO 80309 \\
$^5$Columbia University; New York, NY 10027 \\
$^6$Fermi National Accelerator Laboratory; Batavia, IL 60510 \\
$^7$University of Florida; Gainesville, FL 32611 \\
$^8$Indiana University; Bloomington, IN 47405 \\
$^9$Los Alamos National Laboratory; Los Alamos, NM 87545 \\
$^{10}$Louisiana State University; Baton Rouge, LA 70803 \\
$^{11}$Massachusetts Institute of Technology; Cambridge, MA 02139 \\
$^{12}$Instituto de Ciencias Nucleares, Universidad Nacional Aut\'onoma de M\'exico, D.F. 04510, M\'exico \\
$^{13}$University of Michigan; Ann Arbor, MI 48109 \\
$^{14}$Saint Mary's University of Minnesota; Winona, MN 55987 \\
$^{15}$Yale University; New Haven, CT 06520\\
$^{16}$Center for Neutrino Physics, Virginia Tech; Blacksburg, VA 24061\\
}

\begin{abstract}

The largest sample ever recorded of $\numub$ charged-current quasielastic (CCQE, $\numub + p \to \mup + n$) candidate events is used to produce the minimally model-dependent, flux-integrated double-differential cross section $\frac{d^{2}\sigma}{dT_\mu d\uz}$ for $\numub$ CCQE for a mineral oil target.  This measurement exploits the large statistics of the MiniBooNE antineutrino-mode sample and provides the most complete information of this process to date.  In order to facilitate historical comparisons, the flux-unfolded total cross section $\sigma\left( E_\nu \right)$ and single-differential cross section $\frac{d\sigma}{d\qsq}$ on both mineral oil and on carbon are also reported.  The observed cross section is somewhat higher than the predicted cross section from a model assuming independently-acting nucleons in carbon with canonical form factor values.  The shape of the data are also discrepant with this model.  These results have implications for intranuclear processes and can help constrain signal and background processes for future neutrino oscillation measurements. 

\end{abstract}

\maketitle
  
\section{\label{sec:Intro} Introduction}  

With the recent determination of the nonzero value of $\theta_{13}$~\cite{dayaBay,reno,doubleCh,t2k,minosTh13}, present and future neutrino oscillation experiments will focus on measurements of the neutrino mass ordering and searches for leptonic CP violation.  To reach discovery-level sensitivity to each of these effects, GeV-range $\nue$ and $\nueb$ appearance must be observed in a long-baseline program with few-percent precision~\cite{nova,t2k2,lbne,memphys,hyperk}.  To facilitate such an ambitious program, the cross section for signal and background $\numu$, $\nue$, $\numub$, and $\nueb$ charged-current processes must be known to high precision.  This goal is commonly met by using a near detector to directly measure the rate and shape of the unoscillated spectrum.  However, if the cross sections are not independently and precisely understood, the extracted information may be model dependent and significantly biased from their true value~\cite{martiniOsc,moselEnuReco}.  In the absence of a near detector, detailed knowledge of the contributing reactions is even more critical to the successful execution of these measurements.  While the experimental and theoretical knowledge of GeV-range neutrino interactions on nuclear targets is improving, the experimental precision of interactions in this range is not better than 10\%~\cite{revMod}.  Of even more concern, as will be discussed in more detail, the fundamental processes contributing to neutrino interactions with nuclear matter are not well understood.  

In general, antineutrino cross sections in the few-GeV region are not as well known as their neutrino counterparts, and in particular there are no charged-current antineutrino cross-section measurements below 1~GeV.  In this work we present the first measurement of the antineutrino charged-current quasielastic (CCQE) double-differential cross section with respect to kinematic properties of the outgoing muon.  These data are obtained using a muon antineutrino beam with mean energy $\langle E_{\bar{\nu}} \rangle$~=~665 MeV and an exposure of 10.1 $\times$ 10$^{20}$ protons on target (POT).  This measurement represents an important step towards reaching the level of knowledge required for next-generation oscillation measurements.

Apart from the valuable constraint these results provide for future experiments seeking to use antineutrino events to measure the few remaining unknown fundamental properties of neutrinos, the interpretation of the data will offer insight into an emerging puzzle.  These results significantly contribute to the body of experimental information that suggest the canonicallyÊused model in neutrino generators of the relativistic Fermi gas~\cite{RFG} (RFG) is insufficient for describing neutrino interactions in nuclear media.  It has been argued elsewhere that the discrepancy may come from inadequate form factors or a combination of the nuclear model and the relevant form factors~\cite{paz}.  The RFG assumes the impulse approximation, under which nucleons housed in dense material are treated as quasifree, independently acting participants subject to a global binding energy and Fermi motion, while the surrounding environment is entirely passive.  In this formalism the interaction is parametrized by a set of tensor, vector, and axial-vector form factors~\cite{LS}.  The vector form factors are measured in electron scattering data~\cite{BBA} while the axial-vector form factor is left to be empirically determined by neutrino experiments and is typically assumed to take a dipole form:

\begin{center}
\begin{equation}
\label{eqn:fa}
F_{A} = \frac{g_{A}}{\left(1+\frac{Q^{2}}{M_{A}^{2}}\right)^{2}}\,\, ,\end{equation}
\par\end{center} 

\noindent where $g_A$ is measured from nuclear beta decay~\cite{PDG}, $Q^2$ is the squared four-momentum transfer and, while constraints exist from pion electroproduction data~\cite{MAMeas}, neutrino experiments usually treat the axial mass $M_A$ as a free parameter.

By measuring the total rate of CCQE interactions and fitting the inferred $Q^2$ distribution, a variety of experiments employing bubble-chamber detectors housing mostly light nuclear targets typically produced consistent measurements of $M_{A}$.  From these data, the averaged value is $M_{A} = 1.026 \pm 0.021$ GeV~\cite{MAMeas,BBBA}.  With the discovery of neutrino oscillations, the use of light nuclear targets for the detection medium became impractical, as the statistics required to make high-precision oscillation measurements are much more easily obtained using dense targets.  With these relatively heavy nuclei and higher-precision detectors, more recent experiments have extracted values of $M_A$ systematically higher than 1.026~GeV~\cite{qePRD,K2KsciFi,MINOS,SB}.  Meanwhile, the modern heavy nuclear target experiment NOMAD has measured a value of $M_{A}$ consistent with the bubble-chamber analyses~\cite{NOMAD}, and preliminary shape results from the MINER$\nu$A experiment seem to also favor $M_A$~$\sim$~1~GeV~\cite{kevinNuInt11}.  

An essential first step to understanding this apparent discrepancy is to recognize the particulars of the model dependence introduced by comparing values of $M_A$ between the many experiments.  Important experimental differences that may contribute to the discrepancy include disparate neutrino spectra, different neutrino detection technologies and the size of the nuclear media employed.  The liberties taken to compare $M_A$ values across these scattering experiments include the dipole form of $F_A$, various expectations of hadronic activity consistent with single-nucleon ejection, and the previously mentioned independent nucleon assumption implicit in both the formalism and in the inference of the $Q^2$ distribution.  A possible reconciliation between the data sets has been proposed through a mechanism resulting in intranuclear correlations of greater strength than previously expected (see Ref.~\cite{qeReview} and references therein).  Such a mechanism is consistent with observations in electron scattering data~\cite{elScat1,elScat2}.  If this process is confirmed for weak interactions via neutrino scattering, its detailed understanding will significantly expand knowledge of intranuclear behavior, and some neutrino oscillation results may need to be revisited~\cite{martiniOsc,moselEnuReco}.  The best chance to definitively resolve this crucial ambiguity lies in the community's ability and willingness to produce and compare model-independent information in both the leptonic and hadronic interaction sectors between experimental data and theoretical calculations.  For this reason, the main result of this work is the double-differential CCQE cross section $\left(\dbldiffl\right)$ on mineral oil, where no assumptions about the underlying process is necessary for its reconstruction.  Regardless of the fundamental interactions contributing to the sample studied, this work reports the first cross-section measurements of $\sim$GeV antineutrinos and thus significantly advances the community's preparedness to search for CP violation with neutrinos. 

This paper is organized as follows: The MiniBooNE experiment is described in Section~\ref{sec:BooNE} while Section~\ref{sec:nuInt} describes the model for neutrino interactions.  The analysis is presented in Section~\ref{sec:xsec}, and the conclusions are summarized in Section~\ref{sec:conc}.  Appendix~\ref{apndx:mumCap} presents a measurement of the $\numu$ charged-current background to the analysis sample, which exploits $\mum$ nuclear capture.  Various model-dependent $\numub$ CCQE cross sections are provided in Appendices~\ref{apndx:modDepCH2} and~\ref{apndx:modDepC}, and Appendix~\ref{apndx:tables} tabulates all cross-section results.

\section{\label{sec:BooNE} The MiniBooNE Experiment}


\subsection{\label{sbsec:mbFlux} Beam line and flux}

\indent MiniBooNE observes an on-axis neutrino flux from the Fermilab Booster neutrino beam line (BNB). A beam of 8.9~GeV/c momentum protons is extracted from the Booster synchrotron in bunches of $5 \times 10^{12}$ protons over 1.6~$\mu$s at an average rate of up to 5 Hz.  A lattice of alternatively focusing and defocusing quadrupole magnets steers the proton spills into a beryllium target 71 cm (1.75 interaction lengths) long.  The protons collide with the target to create a spray of secondary particles.  An aluminum electromagnetic horn surrounding the target is pulsed to coincide with the p-Be collisions, creating a toroidal magnetic field to focus mesons of the desired charge.  For the data used in this analysis, the polarity of the magnetic horn is set such that negatively charged secondary particles are focused while those with positive charge are defocused.  The accepted mesons are allowed to decay in a 50~m long air-filled hall, which terminates at a steel beam dump.  The dominant decay modes of these mesons, mostly pions, produce muon neutrinos and antineutrinos.  

\begin{figure}[h]
\includegraphics[scale=0.46]{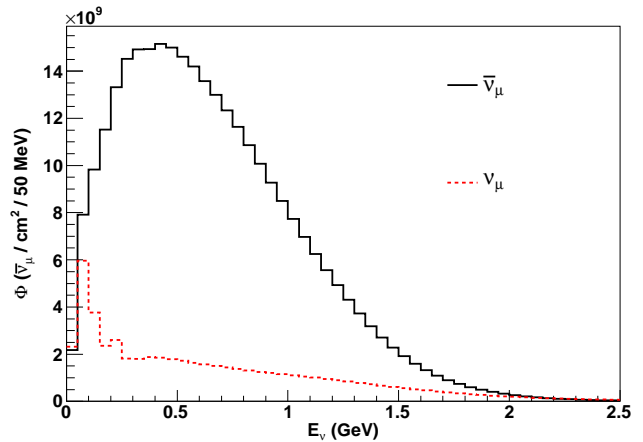} \\
\caption{(Color online) The MiniBooNE $\numub$ and $\numu$ flux predictions for antineutrino-mode for the 10.1$\times$10$^{20}$ POT exposure used in this analysis.  Numerical values for the $\numub$ flux are provided in Table~\ref{tbl:nubflux}.}
\label{fig:nubFlux}
\end{figure}

\begin{figure}[h]
\includegraphics[scale=0.46]{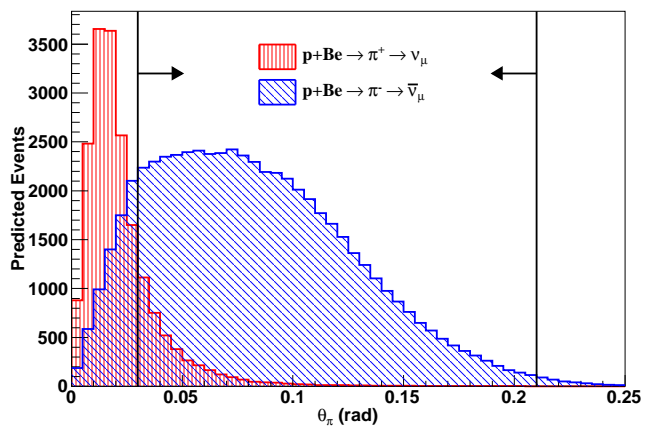} \\
\caption{(Color online) Predicted angular distributions of pions with respect to the incident proton beam ($\theta_{\pi}$) producing $\numu$ and $\numub$ events in the MiniBooNE detector in the antineutrino-mode beam configuration.  The $\numu$/$\numub$ event fraction is significantly larger than the flux fraction due to the respective cross sections.  Distributions are normalized to 10.1 $\times \,\, 10^{20}$ POT, and arrows indicate the region of HARP data~\cite{HARP} constraints.  Figure taken from Ref.~\cite{wsPRD}.}
\label{fig:thPi}
\end{figure}

At MiniBooNE's request, the HARP experiment measured pion production cross sections with a 8.9 GeV/c momentum proton beam on a 5\% interaction length replica MiniBooNE target~\cite{HARP}.  The HARP double-differential cross section in pion energy and angle minimizes the model dependence of the BNB neutrino flux calculation~\cite{mbFlux}. A \textsc{geant}4-based model~\cite{GEANT4} takes these data as input and is used to predict the flux of neutrino and antineutrinos observed by the detector.  The simulation considers proton transport to the target, p-Be interactions in the target including secondary interactions, meson production and their propagation through the magnetic field inside the horn, meson decay, and finally neutrino propagation to the detector.  The uncertainty of primary $\pim$ production at the target is based exclusively on the HARP $\pim$ double-differential cross section~\cite{HARP}.  Though the beryllium target used to collect the HARP data is substantially shorter compared to the MiniBooNE target (5\% vs. 170\% interaction lengths, respectively), the difference in $\pi$ production arising from the thickness between the two targets is calculated to be small.  For the proton energies used by the BNB, roughly 90\% of the neutrino beam is expected to come from the decay of primary $\pi$'s~\cite{sachaRev}, making the MiniBooNE flux prediction minimally dependent on the model for reinteractions in the long target.  The antineutrino-mode beam intersecting the detector is composed of 83.7\% $\numub$, 15.7\% $\numu$, 0.4\% $\nueb$, and 0.2\% $\nue$.  The $\numu$ and $\numub$ flux predictions are presented in Figure~\ref{fig:nubFlux}.  The electron-type neutrinos are irrelevant to this analysis, but as the MiniBooNE detector is unmagnetized, the $\numu$ contribution represents a significant background.  Furthermore, Figure~\ref{fig:thPi} shows that, while the majority of $\numub$'s produced by $\pim$ decay are constrained by the HARP measurement, most of the $\numu$ originating from $\pip$ decay arise from a region not reported by HARP.  In the analysis, the accepted flux of $\numu$ in the antineutrino-mode beam is thus constrained using the observed rate of $\numu$ events in the MiniBooNE detector, as presented in Ref.~\cite{wsPRD} and Appendix~\ref{apndx:mumCap}.  These analyses constrain the knowledge of the $\numu$ flux and the number of neutrino events in the antineutrino-mode sample to less than 15\% for the bulk of the spectrum.   The fractional uncertainty of the $\numub$ flux prediction is around 7\% at the interaction peak, due in roughly equal amounts to errors on $\pim$ production and the model that connects their production to the $\numub$ flux.

Further details of the beam line and flux prediction are given in Ref.~\cite{mbFlux}.

\subsection{\label{sbsec:mbDet} Detector}

The detector is a 12.2 m diameter sphere filled with 818 tons of undoped mineral oil.  The tank is optically segregated into an inner signal region of radius 575~cm and an outer veto shell of 35~cm thickness.  Light produced in the detector is collected by 1520 8-inch Hamamatsu photomultiplier tubes (PMTs), 1280 of which face into the signal region (11.3\% coverage) while 240 are inside the outer shell.  Low activity in the veto region is required in physics analyses to ensure containment of charged particles produced by beam-induced neutrinos while also eliminating contamination from charged particles entering the tank. 

\indent Kept at $\sim$ $20\,^{\circ}\,\mathrm{C}$, the mineral oil has a density of 0.845~g/cm$^{3}$ with an index of refraction of 1.47.  Under these conditions, charged particles with velocity $\beta\, > 0.68$ produce Cherenkov radiation.  Lepton particle identification and reconstruction is principally obtained through the pattern and timing of this prompt Cherenkov light.  The PMTs have a quantum efficiency of $\sim$ 20\% and a timing resolution of $\sim$2~ns, and the prompt Cherenkov component is easily separable from the delayed scintillation light present due to impurities in the oil.  Four dispersion flasks at various locations in the detector are used to illuminate the signal-region PMTs with light from a pulsed laser.  The laser data provide a calibration of PMT responses and allows an {\it in situ} measurement of light scattering properties over time.  Throughout more than ten years of MiniBooNE running, the observed energy scale has been stable to within 1\%.

\indent PMT charge and time information is collected for a total of 19.2~$\mu$s beginning $\sim$5~$\mu$s before the 1.6~$\mu$s long proton spill from the BNB.  Cosmic ray muons stopped in the signal region prior to the start of the DAQ window may decay in time with the BNB spill, so PMT activity 5~$\mu$s before proton delivery is monitored and used to minimize this contamination.  Activity is recorded subsequent to the beam window for more than 10~$\mu$s to observe electrons from the at-rest decay of muons (hereafter referred to as ``Michel" electrons) produced directly or indirectly through the primary neutrino interaction.   

\indent The detector response to muons is calibrated using a dedicated system that independently measures the energy and direction of cosmic ray muons up to 800 MeV.  A scintillator hodoscope directly above the detector and seven scintillator cubes at various depths within the detector are used to track these particles.  Each cube is connected by optical fiber to a PMT for readout.  Signals generated in the hodoscope and PMTs consistent with a muon stopping in a scintillator cube afford a direct calibration of the detector response to the range of muon kinematics most important to this analysis.  These signals are used to verify muon reconstruction algorithms.  Full reconstruction details are available in Ref.~\cite{rbPatterReco}, while the detector is described further in Ref.~\cite{mbDet}. 

\section{\label{sec:nuInt} Predicted neutrino interactions}

MiniBooNE uses the \textsc{nuance}~\cite{nuance} event generator to simulate neutrino interactions.  \textsc{nuance} includes a comprehensive neutrino and antineutrino cross-section model which considers known interactions in the neutrino and antineutrino energy range from $\sim$100~MeV to 1~TeV.  Ninety-nine reactions are modeled separately and combined with nuclear models describing bound nucleon states and final-state interactions to predict event rates and kinematics.

Bound nucleons in the detector medium are described by the RFG~\cite{RFG}.  This assumes the nucleons to be independent and quasifree.  Also specified is a hard cutoff in available struck nucleon energies as dictated by the exclusion principle.  

The neutrino interaction types most relevant to the current analysis are charged-current quasielastic (Section~\ref{sbsec:QExsec}) and single-pion production (Section~\ref{sbsec:piInts}).  The neutrino-induced absolute cross sections for both processes have been measured at MiniBooNE using a flux prediction well determined by HARP data~\cite{HARP}.  These cross-section measurements are utilized in the antineutrino-mode simulation.
More broadly, to minimize the model dependence of the extracted $\numub$ CCQE cross section, each clear opportunity to constrain the backgrounds using MiniBooNE data was exploited.

\subsection{\label{sbsec:QExsec} Charged current quasielastic scattering}

CCQE interactions are the most prevalent channel in MiniBooNE's energy range, and are predicted to account for $\sim$ 40\% of all events. Their simulation in this analysis is chosen based on results from the MiniBooNE $\numu$ and $\numub$ CCQE data sets.  The formalism is described by a relativistic Fermi gas model~\cite{RFG}, and with a few empirical parameter adjustments, this model adequately reproduces the kinematics of both CCQE data sets~\cite{qePRD,JG_nuInt11}.  Through the procedure to correct for detector resolution effects, this choice only mildly affects the shape of the extracted true muon kinematics, while the normalization of the distribution is entirely unaffected.  It will be shown later that this effect is negligible compared to other systematic uncertainties.

The vector and tensor components of the interaction are constrained by data from electron scattering experiments and a nondipole form is taken based on the results of Ref.~\cite{BBA}.  As shown in Equation~\ref{eqn:fa}, the axial-vector form factor assumes a dipole form and contains the empirical ``axial mass" parameter $M_{A}$.  In this analysis, the value of $M_A$ is chosen based on results from neutrino interactions.

As $\numu$ CCQE interactions exclusively interact with nucleons bound in carbon, $\maeffc$ = 1.35~GeV together with a Pauli blocking adjustment, $\kappa$ = 1.007 is sufficient to describe the kinematics of such events based on a fit to the MiniBooNE data~\cite{qePRD}.  The parameter $\kappa$ scales the lowest allowed outgoing nucleon energy for interactions with carbon: $E_{\textrm{low}} = \kappa \left(\sqrt{k_F^2 + M^2} - \omega + E_B \right)$, where $k_F, M, \omega$, and $E_B$ are the Fermi momentum, nucleon mass, energy transfer, and binding energy, respectively.  With the kinematics of $\numu$ CCQE interactions characterized by this adjusted prediction, the total cross section for the simulated process is subsequently corrected to the observed normalization in data.  In this way, the details of the observed $\numu$ CCQE data are reproduced in the present simulation for this process.  


The MiniBooNE mineral oil is composed of C$_{n}$H$_{2n+2}$, $n \sim 20$, and so CCQE scattering off of both bound and quasifree protons are accessible to $\numub$'s.  For the hydrogen scattering component,   $\maeffh$ = 1.02~GeV is chosen based on the body of experimental results for the CCQE process incident on light nuclear targets~\cite{MAMeas,BBBA}.  For interactions with protons bound in a carbon nucleus, the binding energy (Fermi momentum) is set to 30~MeV (220~MeV) based on electron scattering data for the QE process~\cite{elQE}.  As electron QE scattering probes all nucleons while QE interactions with neutrinos and antineutrinos are sensitive to a specific nucleon type, the binding energy determined from electron scattering data is adjusted based on estimates of Coulomb and isospin effects~\cite{tkThesis}.  Along with the same CCQE model parameters measured in the $\numu$ data of $\maeffc$ = 1.35~GeV and $\kappa$ = 1.007, these choices are adopted for $\numub$ CCQE interactions on carbon.  This choice is made exclusively due to observed agreement between this model and the MiniBooNE $\numub$ CCQE data~\cite{JG_nuInt11}.  Note that, due to the axial-vector interference term in the formalism, the kinematics of $\numub$ CCQE features a softer momentum-transfer spectrum and so, in the RFG, the same value of $\kappa$ has a larger effect on $\numub$ CCQE compared to $\numu$ CCQE.  More importantly, it will be shown later that the extracted $\numub$ CCQE double-differential cross section is only negligibly affected by these choices.

The superscript ``eff", short for ``effective", on $M_{A}$ is used throughout this work to allow for the possibility that nuclear effects are responsible for the apparent discrepancy between the light target results and those from the more recent experiments using dense nuclear material.  As discussed in the Introduction, this is also motivated by theoretical work that predicts an extra class of events whose contribution to the CCQE sample in Cherenkov detectors, such as MiniBooNE, enters due to the lack of requirement on hadronic activity~\cite{qeReview}.  The letter following the ``eff" identifies the relevant nucleon target in MiniBooNE's hydrocarbon medium: H for the quasifree hydrogen targets and C for those bound in carbon.   

\subsection{\label{sbsec:piInts} Pion production}

The majority of single-pion production ($\nu_l + N \to \l + \pi + N'$) events at MiniBooNE energies are mediated by baryonic resonances.  The formalism to describe these events is taken from the Rein-Sehgal model~\cite{R-S}, where the relativistic harmonic oscillator quark model is assumed~\cite{Feyn}.  The production of $\De$(1232) is dominant in the energy range spanned by MiniBooNE, but 17 other and higher-mass resonances are also considered. 

The charged-current single-pion channels for $\numu$ ($\numu + N \to \mum + \pip + N$, ``CC1$\pip$") and $\numub$ ($\numub + N \to \mup + \pim + N$, ``CC1$\pim$") dominate the pion-producing interactions  contributing to the $\numub$ CCQE sample.  The CC1$\pip$ events enter from the $\numu$ content of the beam.  The CC1$\pim$ background results from $\numub$ interactions, and their presence in the CCQE sample is mostly due to stopped $\pim$ capture in the nuclear medium.  Stopped $\pim$ capture in the presence of carbon is $\sim$ 100\%, so they are not separable from the $\numub$ CCQE sample through observation of $\pim$ decay.  In the current analysis, the Rein-Sehgal prediction for both classes of events is adjusted to reproduce the kinematic distributions measured in MiniBooNE neutrino-mode CC1$\pip$ data~\cite{qePRD,CCpip}.

\subsection{\label{sbsec:finState} Final-state interactions}

Subsequent to a neutrino interaction involving a nucleon bound in carbon, $\nuance$ propagates the outgoing hadrons including nucleons, mesons and baryonic resonances, and simulates their reinteraction as they exit the nucleus.  The initial interaction model employs the impulse approximation which assumes an instantaneous exchange with independent nucleons.  Subsequent to the initial neutrino or antineutrino interaction, particles produced inside the nucleus are propagated stepwise in 0.3 fm increments until they emerge from the $\sim$ 2.5 fm radius sphere.  Intermittently, the probability for hadronic reinteraction is calculated using a radially dependent nucleon density distribution~\cite{nuclDist} along with external $\pi - N, N - N$ cross-section measurements~\cite{piN}.  For $\De$ reinteractions ($\De + N \to N + N$), an energy-independent probability of 20\% (10\%) is taken for $\De^{+} + N$, $\De^{0} + N$ ($\De^{++} + N, \De^{-} + N$) based on K2K data~\cite{K2Kpr} and is assigned 100\% uncertainty.  The dominant final-state interactions affecting this analysis are pion charge exchange ($\pi^{\pm} + X \leftrightarrow \piz + X^{'}$) and absorption ($\pi^{\pm} + X \to X^{'}$).

\section{\label{sec:xsec} Analysis}

This section describes the extraction of the $\numub$ CCQE double-differential cross section.  It is necessary to first identify the experimental complications that distinguish this measurement from the MiniBooNE $\numu$ CCQE result.


Though the same detector, reconstruction and event selection are used for the $\numu$~\cite{qePRD} and $\numub$ CCQE analyses, subtleties related to the detector material and the different beam configuration result in substantially different sample content in both the signal and background processes.  Due to leading-particle effects at the beryllium target, the mean energy of the $\numub$ flux in antineutrino mode (shown in Figure~\ref{fig:nubFlux}) is appreciably lower ($\langle E_{\nub} \rangle$~=~665 MeV) compared to the $\numu$ flux in neutrino mode ($\langle E_{\nu}\rangle~=$~788 MeV).  The content of the two CCQE signal samples is also fundamentally different since $\numub$ CCQE events arise from interactions with protons while $\numu$ CCQE events involve interactions on neutrons.  The hydrocarbon nature of the detection medium provides a mix of bound and quasifree interaction targets for $\numub$ CCQE, while $\numu$ CCQE involves only bound nucleons.  The two interaction types for $\numub$ CCQE are not separable, and so the sum of all $\numub$ CCQE interactions are treated as the signal for this analysis.  However, as historical data on mostly light targets are adequately described with $M_A \sim$ 1~GeV, results evoking this model to subtract the quasifree $\numub$ CCQE content are given in Appendix~\ref{apndx:modDepCH2}.

Backgrounds in this analysis also offer unique complications, as mentioned in Section~\ref{sbsec:mbFlux} and expanded in the next section.  Broadly, the analysis sample is formed with a simple selection that requires the prompt muon be contained in the detector and that its decay is observed.  The dominant backgrounds with this selection are $\numu$ CC and $\numub$ CC1$\pim$ interactions.  The $\numu$ CCQE contribution is indistinguishable event by event from $\numub$ CCQE; however statistical measurements of their overall rate and shape discussed in Section~\ref{sbsec:cons} constrain the knowledge of this background to $\sim$ 15\%.  The CC1$\pim$ contamination enters from the capture process on carbon nuclei and is known less precisely, as it is not separable in the data from $\numub$ CCQE.  Furthermore, there are no measurements of this process in external data sets at the MiniBooNE energy range.  Due to the size of the $\numu$ CCQE and single-pion backgrounds, the signal purity is only 61\% in this work, compared to 77\% for the $\numu$ CCQE analysis.  Multiple dedicated analyses and comparisons were necessary to reduce the uncertainty on these processes to a manageable level, and as a result, the final uncertainty on the extracted $\numub$ cross sections are dominated by the level of $\numub$ flux uncertainty.

\subsection{\label{sbsec:cons} Constraints on background processes}

The largest background in the $\numub$ CCQE sample is the $\numu$ contamination.  Moreover, as Figure~\ref{fig:thPi} shows, the majority of $\pip$ particles contributing to the beam are produced at small angles with respect to the incoming protons (and so are affected less by the magnetic field) and thus their contribution to the antineutrino-mode beam is mostly unconstrained by the HARP hadroproduction data.  As MiniBooNE is nonmagnetized, this motivated a dedicated study of the $\numu$ beam content using statistical methods.  Three techniques, described in detail in Ref.~\cite{wsPRD} and Appendix~\ref{apndx:mumCap}, were used to measure this crucial background for the MiniBooNE data.  Two of these measurements are largely model independent, and the final fractional uncertainty on the $\numu$ contribution to the antineutrino-mode beam is $\sim$15\% for the bulk of the observed spectrum.  These analyses are the first of their kind and their uncertainty has reduced the error on the $\numub$ CCQE cross section due to $\numu$ interactions to a subdominant uncertainty. 

\begin{figure}[h]
\includegraphics[scale=0.46]{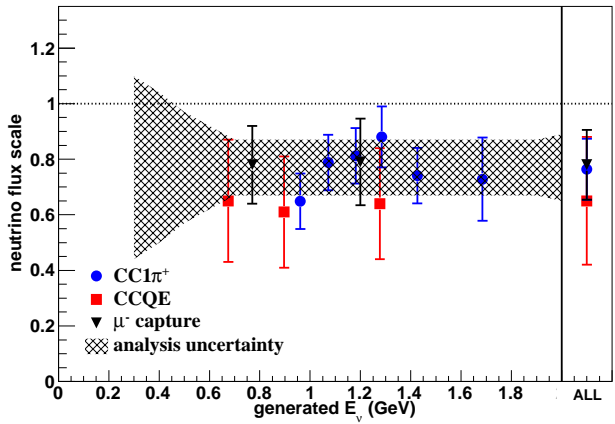} \\
\caption{(Color online) Summary of the results from three techniques used to measure the $\numu$ flux in the antineutrino-mode beam.  The ``$\mum$ capture" analysis is described in Appendix~\ref{apndx:mumCap} and the others in Ref.~\cite{wsPRD}.  The measurements are given relative to an extrapolation of HARP data into a region where no relevant hadroproduction data exist (shown as a dotted line at unity).  Due to dependence on assumptions of $\numub$ CCQE scattering, results from the ``CCQE" determination of the $\numu$ flux are not used in the $\numub$ CCQE cross-section analysis.}
\label{fig:wsSummary}
\end{figure}

The three measurements of the $\numu$ contribution to the antineutrino-mode data exploit various differences between charged-current $\numu$ and $\numub$ processes to statistically measure their respective contributions.  Broadly, these measurements are executed by performing rate analyses on samples with the $\numu$ and $\numub$ content statistically separated.  These techniques include the use of $\mum$ nuclear capture, $\pim$ nuclear capture, and angular differences between $\numu$ and $\numub$ in CCQE interactions.  The analysis based on $\mum$ capture is described in Appendix~\ref{apndx:mumCap} and the other analyses are presented in Ref.~\cite{wsPRD}.  The $\mum$ capture analysis exploits the $\sim$ 8\% of $\numu$-induced CC interactions on carbon that do not lead to a decay electron, while nuclear capture of $\pim$ also affords sensitivity to the $\numu$ beam content.  The second most prevalent interaction type in the MiniBooNE detector is CC single-pion production, which produces a $\pip$ in the case of $\numu$ scattering and a $\pim$ for $\numub$ reactions.  As almost all stopped $\pim$ are absorbed in the hydrocarbon medium~\cite{pimCap}, the sample consisting of a single muon and two decay electrons (one each from the prompt muon and the pion decay chain) is predominantly due to $\numu$ events.  Finally, the observed angular distribution of CCQE events is fit to a combination of $\numu$ and $\numub$ events, where $\numub$ interactions are predicted to be much more forward going with respect to the beam direction.  This last analysis is dependent on details of the RFG prediction for $\numub$ CCQE scattering, and \emph{its results are not used in the subtraction of the $\numu$ background}.  In the future, where CCQE and CCQE-like interactions should be better understood, this technique could provide a valuable constraint.  The results from these analyses are summarized in Figure~\ref{fig:wsSummary}, where the nominal and highly uncertain prediction of the $\numu$ flux in the antineutrino-mode beam appears to be roughly 20\% high in normalization, while the energy dependence seems to be well modeled.  Based on the results from the $\mum$ and $\pim$ nuclear capture analyses, the $\numu$ flux in the antineutrino-mode beam is corrected by a scale of 0.77 with an uncertainty of 0.10.  These values are obtained using a method for combining correlated measurements~\cite{corrMeas} with an estimated correlation coefficient of 0.5 based on the common dependence of the HARP $\pip$ production data in the CC1$\pip$ and $\mum$ capture analyses.  To recognize possible spectral dependencies in these data, the uncertainty of 0.10 is increased outside the regions directly constrained.  This increased uncertainty is particularly important at lower energies, where much of the $\numu$ flux originates in the decay of $\pip$ produced in regions that {\it are} constrained by the HARP measurements.  The uncertainty on the $\numu$ subtraction in the calculation of the $\numub$ double-differential cross section $\dbldiffl$ is shown in Figure~\ref{fig:wsSummary}.  Note these corrections calibrate the primary $\pip$ production cross section in $p+\textrm{Be}$ interactions contributing to the antineutrino-mode beam.  Other systematic effects, such as the modeling of the magnetic field and secondary interactions in the target allow energy-dependent shifts and are evaluated and included in the analysis separately.

The measurements summarized in Figure~\ref{fig:wsSummary} calibrate the simulated $\pip$ production at the beryllium target to the level that the cross sections for the $\numu$ processes contributing to the analysis samples are known.  The most important interactions are the $\numu$ CCQE and CC1$\pip$ processes measured in the MiniBooNE neutrino-mode exposure~\cite{qePRD,CCpip}.  Due to the disparate $\pip$ acceptance to the beam, the $\numu$ flux spectrum in neutrino mode is much harder in energy compared to the $\numu$'s in antineutrino mode.  See Figure~2 of Ref.~\cite{wsPRD}. However, as suggested by Figure~\ref{fig:tmuEff}, high-energy neutrinos are largely rejected by the analysis requirement of contained muons, and the {\it accepted} $\numu$ spectrum between neutrino and antineutrino run modes is nearly identical.  This shared $\numu$ spectrum allows the cross sections extracted from the neutrino-mode data to be directly applied to the antineutrino-mode simulation without relying on knowledge of the relationship between muon kinematics and incident neutrino energy.  As discussed in Refs.~\cite{martiniOsc} and~\cite{moselEnuReco}, this connection [Equation~\ref{eqn:EnuQE}] may be unreliable in the presence of background interactions that originate from intranuclear processes.

\begin{figure}[h]
\includegraphics[scale=0.46]{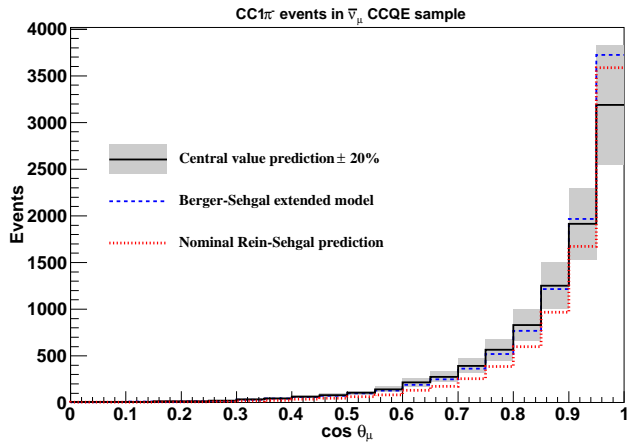} \\
\caption{(Color online) Three calculations for the CC1$\pim$ background contribution to the $\numub$ CCQE sample as a function of $\uz$, where $\theta_\mu$ is the angle of the muon relative to the neutrino direction.  The ``central-value" distribution corresponds to the nominal Rein-Sehgal~\cite{R-S} prediction for CC1$\pim$ events in MiniBooNE constrained by the observed kinematics in the neutrino-mode $\numu$ CC1$\pip$ sample.  This agrees well with a more recent calculation (``Berger-Sehgal extended model") that is based on an improved version of the Rein-Sehgal model.  For comparison, the bare Rein-Sehgal prediction for CC1$\pim$ events is also shown.  Distributions are normalized to 10.1 $\times$ 10$^{20}$ POT.}
\label{fig:ccpim}
\end{figure}

Charged-current single $\pim$ production constitutes the second-largest background to this analysis, accounting for $\sim$ 15\% of the sample.  These events enter through a different and experimentally disadvantageous mechanism compared to the analogous process for the $\numu$ CCQE sample.  Single-pion events induced by $\numu$ typically give rise to Michel electrons through the decay chain $\pip \to \mup \to \elp$ of stopped pions, which can be observed and used to reject these events.  However, an appreciable number of $\pip$ are destroyed in flight through the nuclear absorption process ($\pip + X \to X'$).  In  contrast, almost all single-pion events from $\numub$ interactions enter the $\numub$ CCQE sample since $\pim$ that are not absorbed in flight stop and undergo nuclear capture with $\sim$100\% efficiency.  While this fortuitously allows for the CC1$\pip$-based $\numu$ flux measurement, this also implies CC1$\pim$ events are not separable from the $\numub$ CCQE sample.  This is in contrast to the MiniBooNE $\numu$ CCQE analysis, where the single-pion events tagged through the observation of an additional Michel allows a direct constraint of the rate and kinematics of CC1$\pip$ events.  A correction was thus measured and applied to the background prediction for the $\numu$ CCQE sample~\cite{qePRD}.  This constraint is applied by correcting the CC1$\pip$ events according to the observed reconstructed momentum transfer.  The correction is shown in Fig. 7(b) of Ref.~\cite{qePRD}.  In the absence of such a measurement for CC1$\pim$ interactions, the constraint obtained in neutrino mode for $\numu$ CC1$\pip$ is applied to the CC1$\pim$ Rein-Sehgal prediction described in Section~\ref{sbsec:piInts}.  Figure~\ref{fig:ccpim} shows this prediction agrees well with an external calculation~\cite{jarekProc} for such events.  This alternate model is implemented in $\nuance$ and is based on extensions of the Rein-Sehgal model~\cite{bergerSehgal,kuzmin1,kuzmin2}.  This updated calculation includes muon mass terms and a modified vector form factor to yield better agreement with world pion production data~\cite{sobczyk}.  Consistency between these two predictions for CC1$\pim$ production, along with the level of agreement between the extended Rein-Sehgal calculation and the MiniBooNE CC1$\pip$ data (shown in Ref.~\cite{jarekProc}) suggests an uncertainty of 20\% is sufficient for the CC1$\pim$ background.  Future tests of the accuracy of this prediction may be made through comparisons to the subtracted CC1$\pim$ background, as given in Appendix~\ref{apndx:tables}.

Based on results from the neutrino-mode $\numu$ CC1$\piz$ analysis~\cite{mbCCpiz}, the small contribution from CC1$\piz$ events induced by both $\numu$ and $\numub$ are increased by a factor of 1.6 relative to the $\nuance$ prediction.  The generated predictions for all other interactions, accounting for $<$3\% of the sample, are not adjusted.

\subsection{\label{sbsec:sel} Reconstruction and analysis sample}

The identification of $\numub$ CCQE candidate events relies solely on the observation of a single muon and no final-state $\pip$.  Muon kinematics are obtained by the pattern, timing, and total charge of prompt Cherenkov radiation collected by PMTs.  A likelihood function operating under a muon hypothesis is compared to the topology and timing of the observed PMT hits.  This likelihood function predicts hit patterns and timing based on the interaction vertex and the momentum four-vector of the muon.  The likelihood function simultaneously varies these seven parameters while comparing to the observed PMT hits.  The parameters from the maximized likelihood function yield the reconstructed muon kinematics.  Integrated over the spectrum of observed muons, the resolution of this reconstruction for muon energy (angle) is roughly 8\% (2 degrees)~\cite{rbPatterReco}.   The direct and high-resolution observation of muon properties motivates the choice to present the $\numub$ CCQE cross section as a function of muon kinematics as the main result of this work, while the large statistics of the data set analyzed yield sensitivity to previously unprobed regions of the interaction.

As in the $\numu$ CCQE work, no requirement is made on hadronic activity.  This is an important distinction from the CCQE definitions used by other experiments~\cite{SB,NOMAD}, where a single proton track may be required for $\numu$ CCQE selection.  However, note that in the case of $\numub$ CCQE scattering, where a single ejected neutron is expected, the experimental definition used by tracking detectors is largely based on a single muon track.  Therefore, in general, the selection used by tracking detectors and Cherenkov-based measurements such as this one for $\numub$ CCQE follow each other more closely as compared to the $\numu$ case.

\begin{table*}
\caption{\label{tbl:purAndEff} Sample purity and detection efficiency for all $\numub$ CCQE events, which are due to a mix of scattering on bound and quasifree nuclear targets.  Efficiencies are normalized to events with a generated radius $r <$ 550~cm.}
\begin{ruledtabular}
\begin{tabular}{cccc}
\multirow{2}{*}{Cut \#} & \multirow{2}{*}{Description} & Purity & Efficiency \\ 
 &  & (\%) & (\%) \\
\hline
0 & No cuts & 32.3 & 100 \\
1 & Veto hits \textless\, 6, all subevents & 27.6 & 50.8 \\
2 & First subevent: in beam window $4000<T(\mathrm{ns})<7000$ & 27.7 & 50.3 \\
3 & First subevent: muon kinetic energy $\tmu$ \textgreater\, 200 MeV & 36.9 & 44.0 \\
4 & Only two subevents & 48.4 & 38.8\\
\vspace{2mm}
5 & First subevent: reconstructed vertex radius R $<$ 500 cm & 49.2 & 32.6 \\
\multirow{2}{*}{6} & Distance between subevent reconstructed vertices \textgreater\, 500 cm/GeV $\times\,\tmu$ - 100 cm & \multirow{2}{*}{54.3} & \multirow{2}{*}{30.6} \\
\vspace{2mm}
& Distance between subevent reconstructed vertices \textgreater\, 100 cm & & \\
7 & First subevent: log-likelihood ($\mu$ / $e$) \textgreater\, 0 & 61.0 & 29.5 \\
\end{tabular}
\end{ruledtabular}
\end{table*}

\begin{figure}[h]
\includegraphics[scale=0.46]{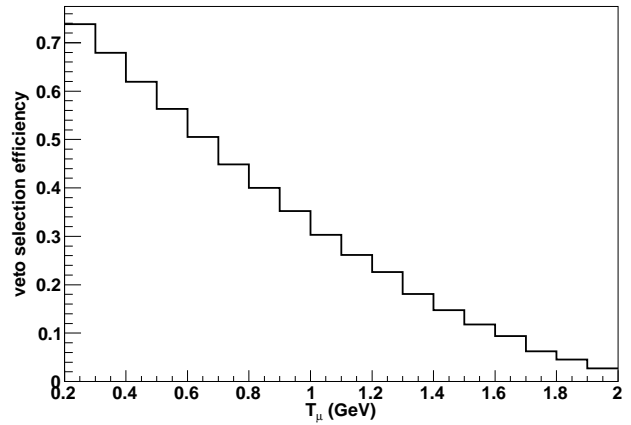} \\
\caption{Selection efficiency as a function of generated muon kinetic energy for $\numub$ CCQE events passing the veto requirement.  Higher-energy muons are less likely to stop in the inner region and are removed by this selection.}
\label{fig:tmuEff}
\end{figure}

In MiniBooNE, final-state neutrons lead to low-energy scintillation light primarily through elastic scattering with the quasifree protons in the hydrogen content of the oil.  The prompt PMT signals that define the analysis sample are dominated by Cherenkov light, and so the delayed scintillation light caused by neutron interactions have a negligible effect on the acceptance of $\numub$ CCQE events.


The event selection is identical to that used in the MiniBooNE $\numu$ CCQE analysis~\cite{qePRD}.  Table~\ref{tbl:purAndEff} provides cumulative purity and efficiency values for the selected sample.  Notice the requirement of low veto activity immediately halves the collection efficiency of $\numub$ CCQE interactions.  As shown in Figure~\ref{fig:tmuEff}, this is primarily due to the rejection of high-energy muons not fully contained within the inner detector region.
Sample selection is based on requirements of temporally correlated collections of PMT activity (or PMT ``hits") referred to as ``subevents".  A hit is any PMT pulse passing the discriminator threshold of $\sim$ 0.1 photoelectrons, and a cluster of at least 10 hits within a 200~ns window with individual hit times less than 10 ns apart defines a subevent.  Two or fewer spacings between 10 - 20~ns among individual hit times are also allowed.  The primary requirement to identify $\numub$ CCQE events is two and only two subevents, due dominantly to Cherenkov light from the prompt muon and its decay positron: 

\beq
\begin{array}{cccl}
1: & \numub + p  & \to &  \mup + n \\ 
2: &            &     &  \hookrightarrow \elp + \nue + \numub 
\end{array} 
\eeq

\begin{figure}[h]
\includegraphics[scale=0.46]{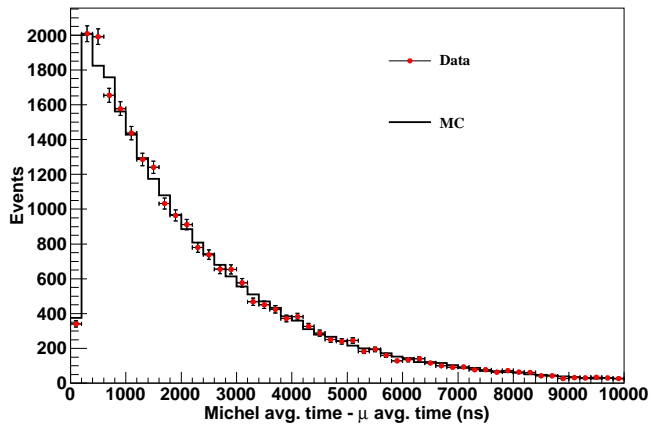} \\
\caption{(Color online) Time separation in the average PMT hit time between the two subevents in the CCQE sample.  The deviation from an exponential form at low time differences is due to the subevent separation requirement. Simulation is normalized to data, and statistical errors are shown with data.}
\label{fig:timing}
\end{figure}

The difference in average PMT hit time between the two subevents is given in Figure~\ref{fig:timing} and shows both the characteristic lifetime of muons in the sample and the effect of the subevent definition on CCQE detection for quickly-decaying muons.  The selection criteria are enumerated in Table~\ref{tbl:purAndEff}.  Cut 1 enforces containment of charged particles produced inside the detector while also rejecting incoming charged particles.  Cut 2 requires the muon subevent be correlated with the BNB proton spill.  Cut 3 ensures the first subevent is not a Michel electron and avoids a region of muon energy with relatively poor reconstruction.  Cut 4 eliminates most neutral-current events and rejects most interactions with final-state $\pip$.  Cut 5 further enhances the reliability of the reconstruction by reducing sensitivity to PMT coverage.  Cut 6 ensures the measurements of the muon energy and the subevent vertices are consistent with the production and subsequent decay of a minimum ionizing particle.  This cut rejects many events where the Michel is not associated with the primary muon, mainly CC1$\pip$ and NC1$\pip$ events where the second subevent is a decay positron from the $\pip$ decay chain.  Cut 7 requires that the candidate primary muon is better fit as a muon than as an electron.  This cut reduces the background from most processes, most notably from CC1$\pip$ and CC1$\pim$.

\begin{figure}[h]
\includegraphics[scale=0.51]{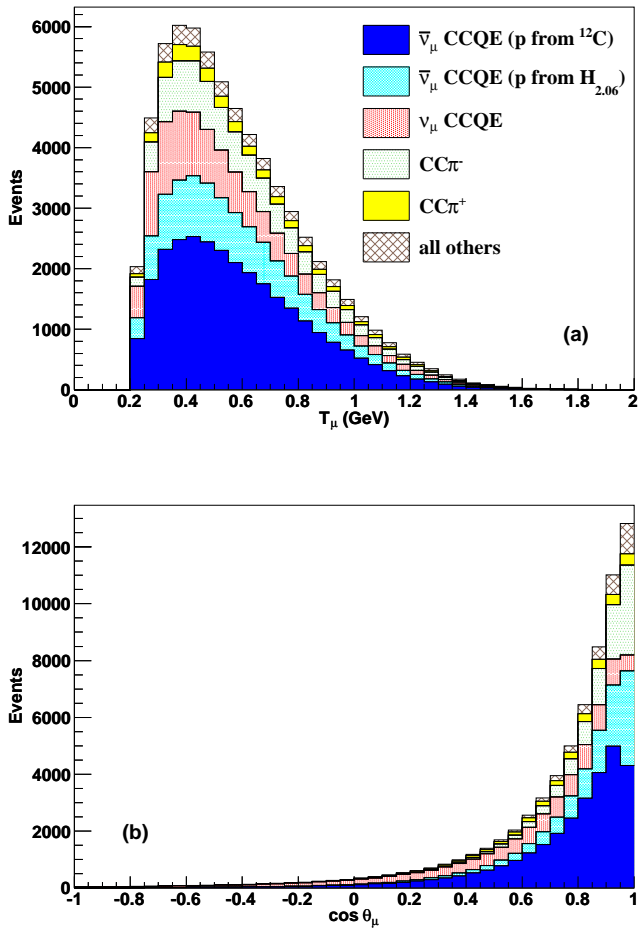} 
\caption{(Color online) Predicted sample composition as a function of reconstructed muon kinematics in the $\numub$ CCQE sample.  The top figure (a) shows the kinetic energy of the muon, while the bottom (b) shows $\uz$, where $\theta_\mu$ is the muon direction relative to the incoming neutrinos.  Distributions are normalized to the exposure of 10.1 $\times$ 10$^{20}$ POT.}
\label{fig:sample}
\end{figure}

For this selection, the neutrino interaction assumptions detailed in Section~\ref{sec:nuInt}, the constrained backgrounds described in Section~\ref{sbsec:cons} and 10.1 $\times$ 10$^{20}$ POT, the sample consists of 71,176 events with $\numub$ CCQE purity (detection efficiency) of 60.3\% (29.5\%).  Table~\ref{tbl:sample} presents a summary of the $\numub$ CCQE sample, while Figure~\ref{fig:sample} shows how these channels contribute to the reconstructed kinematical distributions of the final-state muon.  

\begin{table}
\caption{\label{tbl:sample} Summary of the $\numub$ CCQE sample.  Contributions reflect all adjustments to simulation based on constraints from MiniBooNE data.}
\begin{ruledtabular}
\begin{tabular}{cc}
integrated POT							& 10.1 $\times$ 10$^{20}$ \\
energy-integrated $\numub$ flux			& 2.93 $\times$ 10$^{11}$ $\numub$ / cm$^2$\\
$\numub$ CCQE candidate events 			& 71176 \\
$\numub$ CCQE efficiency ($R < 550$ cm)	& 29.5\% \\
\hline
Interaction channel & Contribution (\%) \\
\hline
$\numub + p$ $\rightarrow \mup + n$ (bound $p$) & 43.2 \\
$\numub + p$ $\rightarrow \mup + n$ (quasifree $p$) & 17.1 \\
$\numu + n \rightarrow \mum + p$ & 16.6 \\
$\numub + N \rightarrow \mup + N + \pim$ (resonant) & 10.4 \\
$\numu + N \rightarrow \mum + N + \pip$ (resonant) & 3.8 \\
$\numub + A \rightarrow \mup + A + \pim$ (coherent) & 3.3 \\
$\numub + N \rightarrow \mup + N + \piz$ & 2.8 \\
\dotfill & \dotfill \\
$\numub + p \rightarrow \mup + \Lambda^{0}$ & \multirow{3}{*}{2.0} \\
$\numub + n \rightarrow \mup + \Sigma^{-}$ & \\
$\numub + p \rightarrow \mup + \Sigma^{0}$ & \\
\dotfill & \dotfill \\
All others & 0.8 \\
\end{tabular}
\end{ruledtabular}
\end{table}

\subsection{\label{sbsec:xsecCal} Cross-section calculation and uncertainties}

The flux-integrated double-differential cross section per nucleon in the $i^{\textrm{th}}$ bin is given by:

\begin{center}
\begin{equation}
\label{eqn:dblDiffl}
\left(\frac{d^2\sigma}{d\tmu\,d\left[\uz\right]}\right)_i = \frac{\sum_j U_{ij}\left(d_j - b_j\right)}{(\Delta \tmu)_i\,(\Delta\left[\uz\right])_i \,\epsilon_i \,\Phi \,N}\,\,\,, \end{equation}
\par\end{center} 

\noindent where $d_j$ refers to data, $b_j$ the background, $U_{ij}$ is an unfolding matrix connecting the reconstructed variable index $j$ to the true index $i$, $\epsilon_i$ is the detection efficiency, $\De \tmu$ and $\De \left[\uz\right]$ the respective bin widths, $\Phi$ the integrated $\numub$ exposure, and $N$ the number of proton targets in the volume studied.  

The unfolding matrix $U_{ij}$ is based on the Bayesian method proposed in Ref.~\cite{unsmear} to account for reconstruction biases.  The high-sensitivity resolution of the reconstruction used to identify muon kinematics leads to only mild corrections.  However, this procedure does introduce some dependence on the generated muon kinematics of $\numub$ CCQE interactions.  This bias is evaluated by unfolding the data with 100 different versions of $U_{ij}$ generated using a conservative range of CCQE model parameters.  The bias introduced by the Bayesian unfolding method for the cross sections reported here is found to be negligible.  Meanwhile, a particular strength of this cross-section configuration is that this unfolding matrix is entirely independent of assumptions regarding the underlying interaction.  This is in contrast to, for example, the total cross section $\sigma(E_\nu)$ computed with only observations of muon kinematics.

It is important to note that by directly subtracting the background from data in the reconstructed distribution, this cross-section extraction procedure is entirely independent of the normalization of the generated signal $\numub$ CCQE processes.  That is, though the RFG with a large value for the effective axial mass is assumed by simulation, the extracted cross section is not affected by this choice.  

Systematic uncertainties are evaluated by forming an error matrix that propagates correlated uncertainties on parameters and processes that affect $\numub$ CCQE interactions onto the calculated cross section.  The covariance matrix is constructed by first forming a distribution of weights corresponding to simulated excursions set by Gaussian variations of parameters and measurements within their associated error.  These weights are then used to recalculate the double-differential cross section in Eqn.~\ref{eqn:dblDiffl}, replacing the central-value MC valuations with the excursion values for terms appropriate to each systematic uncertainty.  The difference of these alternate cross-section calculations compared to the ``best guess" distribution forms the covariance matrix:

\begin{equation}
\begin{array}{l}
M_{ij} = \frac{1}{K}\sum\limits_{s = 1}^{K} (N_i^{s} - N_i^{CV}) \times (N_j^{s} - N_j^{CV}). \\   
\end{array} 
\label{eqn:ErrMat}
\end{equation}

\noindent Here $K$ simulation excursions are used, $N^{s}$ is the reweighted cross-section value corresponding to the $s^{th}$ simulation set and $N^{CV}$ represents the simulation central value.  This technique is further described in Ref.~\cite{brNIM}.  For uncertainties on processes with correlated errors, typically $K$ = 100 while $K$ = 1 is sufficient for uncorrelated errors.  Systematic uncertainties requiring correlated errors are the production of $\pim$ in the proton beam target, the connection between $\pim$ production and the focused $\numub$ beam, optical transport in the detector, final-state interactions, and the bias due to the unfolding procedure.  

As mentioned in Section~\ref{sbsec:mbFlux}, the uncertainty on the production of $\numub$ parent $\pim$ at the beryllium target is driven by the HARP data~\cite{HARP} and the absolute $\numub$ flux prediction is minimally dependent on the hadroproduction model.  Subsequent to $\pim$ production, errors on the processes that culminate in the $\numub$ beam include the amount of delivered POT, optics of the primary beam, magnetic focusing, and hadronic interactions in the target and the enclosing horn.  More details on uncertainties of the flux prediction are available in Ref.~\cite{mbFlux}.  Uncertainties on the model for optical transport in the detector are based on both external and {\it in situ} measurements of light attenuation, scintillation strength, and the refractive index of the oil~\cite{OM}.  For this uncertainty, 70 samples generated with variations of 35 parameters that describe the optical model are used to find the uncertainty propagated to the measurement. The most important final-state interactions affecting the composition of the $\numub$ CCQE sample are the pion charge exchange ($\pi^{\pm} + X \leftrightarrow \piz + X^{'}$) and absorption ($\pi^{\pm} + X \to X^{'}$) processes.  The uncertainty on pion charge exchange (absorption) inside the nucleus is set to 30\% (25\%) based on the difference between the $\nuance$ prediction and external data~\cite{piAbsCEXdata}.  The intermedium processes are evaluated separately with 50\% (35\%) fractional uncertainty based on comparisons with the \textsc{gcalor} prediction and the same external data.  The final correlated systematic error evaluates the bias introduced by the Bayesian unfolding procedure, where 100 different matrices $U_{ij}$ are generated within $M_A$~=~1.35~$\pm$~0.35~GeV and $\kappa$~=~1.007~$\pm$~0.007.  The negligible bias found when the data are extracted with these alternate matrices assuming a conservative range of CCQE parameter values assures this cross-section measurement is largely independent of the CCQE interaction model.

Uncertainties described by a single excursion from the simulation central value include errors due to detector PMT response and on background processes not due to final-state interactions.  Large sets of simulation are generated separately to evaluate rate biases due to uncertainties on phototube discriminator threshold and the correlation between pulse time and delivered charge.  The background processes are grouped into three classes: CC1$\pim$ events, those induced by $\numu$, and all non-CCQE, non-CC single-pion events.  Note these groups are not mutually exclusive and all constraints are described in Section~\ref{sbsec:cons}.  Based on consistency of the prediction using an extrapolated CC1$\pi$ constraint with a robust external model for the CC1$\pim$ background, these events are assigned 20\% uncertainty.  All $\numu$ background events are subject to the measured uncertainty shown in Figure~\ref{fig:wsSummary}.  The cross sections for the $\numu$ CCQE and CC1$\pip$ processes are directly measured by MiniBooNE data~\cite{qePRD,CCpip}, and so only their flux is uncertain.  The uncertainty on the small contribution from coherent $\pi$ production is set to 60\%, while the other non-CCQE, non-CC1$\pipm$ processes are assigned 30\% cross-section uncertainty. 

The overall size of these covariance matrices can be expressed with a single number, representing the normalization uncertainty of each error.  Using the sum rule for variances and covariances, the total normalization uncertainty can be thought of as the error on the cross section if the measurement consisted of a single bin: 

\begin{equation}
\delta D_T / D_T = \sqrt{ \sum\limits_{ij}^n M_{ij}} / \sum\limits_i^n D_i\,\,, 
\end{equation}

\noindent where $D_T = \sum\limits_i^n D_i$ represents the double-differential cross-section measurement summed over each kinematic region $i$.  This is also commonly referred to as the uncertainty on the scale of the measurement.  Table~\ref{tbl:normUnc} shows the contributions of various errors to the total normalization uncertainty.

The covariance matrix can also be used to separate the correlated normalization uncertainties from the total error, leaving information related to how much the shape of the observed data may vary within the systematic errors~\cite{tkThesis}.  These uncertainties are identified by first defining a data vector $V$ with entries corresponding to the observed relative normalization of each bin: $V_i = \{ D_1 / D_T, D_2 / D_T, \dotsb, D_n / D_T, D_T \}$.  Notice this vector has dimension $n+1$, where $n$ is the number of bins measured.  The covariance matrix $Q$ for this new vector $V$ involves the Jacobian matrix of partial derivatives $J$ and is given by: 

\begin{equation}
Q_{kl} = \sum\limits_{ij}^{n} J_{ki} M_{ij} J_{lj} = \sum\limits_{ij}^{n} \frac{\partial V_k}{\partial D_i} M_{ij} \frac{\partial V_l}{\partial D_j}\,\,\,.
\end{equation}

\noindent The diagonals of the matrix $Q$ are related to the shape uncertainty in each kinematic bin.  For entries $\{1,2,\dotsb,n\}$,

\begin{eqnarray}
Q_{kk} &=& \frac{1}{D_T^2} \left[M_{kk} - 2\frac{D_k}{D_T}\sum\limits_i^n M_{ik} + \frac{N_k^2}{N_T^2} \sum\limits_{ij}^n M_{ij}\right] \\
       &=& \left(\delta D_{k,\textrm{shape}}\right)^2 \nonumber \end{eqnarray}

As the full covariance matrix $M$ for the double-differential cross section is in principle a four-dimensional object with over 100,000 entries, the combination of the total normalization error and the bin-by-bin shape error is the preferable method to report the complete experimental uncertainty.  This is argued more completely in Ref.~\cite{unsmear}, and Ref.~\cite{janNumuQEfit} provides an example of how to use this information in the context of a fit to these data.

\begin{table}
\caption{\label{tbl:normUnc} Normalization uncertainty for various sources of error for the $\numub$ CCQE cross section on mineral oil.}
\begin{ruledtabular}
\begin{tabular}{cc}
Uncertainty type & Normalization uncertainty (\%) \\
\hline
$\numub$ flux 		& 9.6 \\
Detector 			& 3.9 \\
Unfolding			& 0.5 \\
Statistics			& 0.8 \\
$\numu$ background 	& 3.9 \\
CC1$\pim$ background 	& 4.0 \\
All backgrounds	 	& 6.4 \\
\hline
{\bf Total}			& {\bf 13.0} \\
\end{tabular}
\end{ruledtabular}
\end{table}
 
The main result of this work is the $\numub$ CCQE double-differential cross section on mineral oil.  However, as the majority of the bubble-chamber CCQE analyses using light targets for the interaction medium are adequately described with $M_A \sim$ 1~GeV~\cite{MAMeas,BBBA}, the cross section on carbon only is found by assuming this value to subtract the quasifree hydrogen content of the $\numub$ CCQE data.  This alternate cross section is calculated by including $\numub$ hydrogen CCQE events in the background term $b_j$ in Equation~\ref{eqn:dblDiffl}, while the other terms in the calculation based on the signal definition now are based on only $\numub$ CCQE events involving protons bound in carbon.  Most notably, this reduces the number of interaction targets in the fiducial volume. 

Informed by the results of fits to the light-target CCQE experiments, $M_A^{\textrm{eff,H}}$~=~1.026~$\pm$~0.021~GeV~\cite{MAMeas,BBBA} is assumed and subtracted from the data.  Systematic error due to this background is evaluated with the method described earlier in this section with $K = 100$ throws against the 0.021~GeV uncertainty.  Including this additional error and, more importantly, considering the lower sample purity for this alternate definition of signal events, the fractional normalization uncertainty increases to 17.4\%.   

\subsection{\label{sbsec:results} Results}
 
\begin{figure}[h]
\includegraphics[scale=0.46]{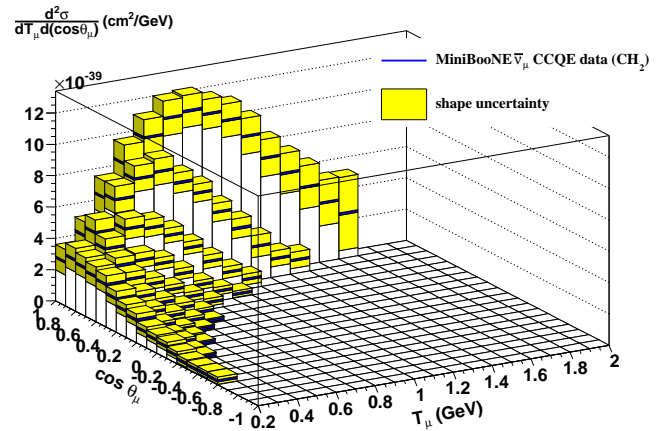} \\
\caption{(Color online) The $\numub$ CCQE per-proton double-differential cross section with shape uncertainty.  The normalization uncertainty of 13.0\% is not shown.  Numerical values for this cross section and its uncertainty are provided in Tables~\ref{tbl:dblDifflCH2} and~\ref{tbl:dblDifflShapeCH2}, respectively.}
\label{fig:twoD}
\end{figure}

The $\numub$ CCQE double-differential flux-integrated cross section on mineral oil is shown with shape uncertainty in Figure~\ref{fig:twoD} and the one-dimensional projections are compared to RFG predictions in Figure~\ref{fig:twoDwComps}.  The configuration with the hydrogen content subtracted is given in Appendix~\ref{apndx:tables} and may be more readily compared to theoretical calculations for $\numub$ CCQE interaction on carbon, such as in Refs.~\cite{Athar,Martini2,Nieves,SuSAnub,Meucci,Bodek}.  Bins in the kinematic region -1~$<~\uz~<~$~+1 and 0.2~$<~T_\mu$~(GeV)~$<~2.0$ are reported if they meet the statistical requirement of at least 25 events in the reconstructed and background-subtracted data term $\left(d_j - b_j\right)$ in Equation~\ref{eqn:dblDiffl}.  If this threshold is not met, no measurement is reported.  As no explicit assumptions about the underlying interaction are necessary to reconstruct muon kinematics, this result is nearly model independent.  Since some background processes are not directly constrained by data, most notably CC1$\pim$, Appendix~\ref{apndx:tables} tabulates the subtracted data.

\begin{figure}[h]
\includegraphics[scale=0.75]{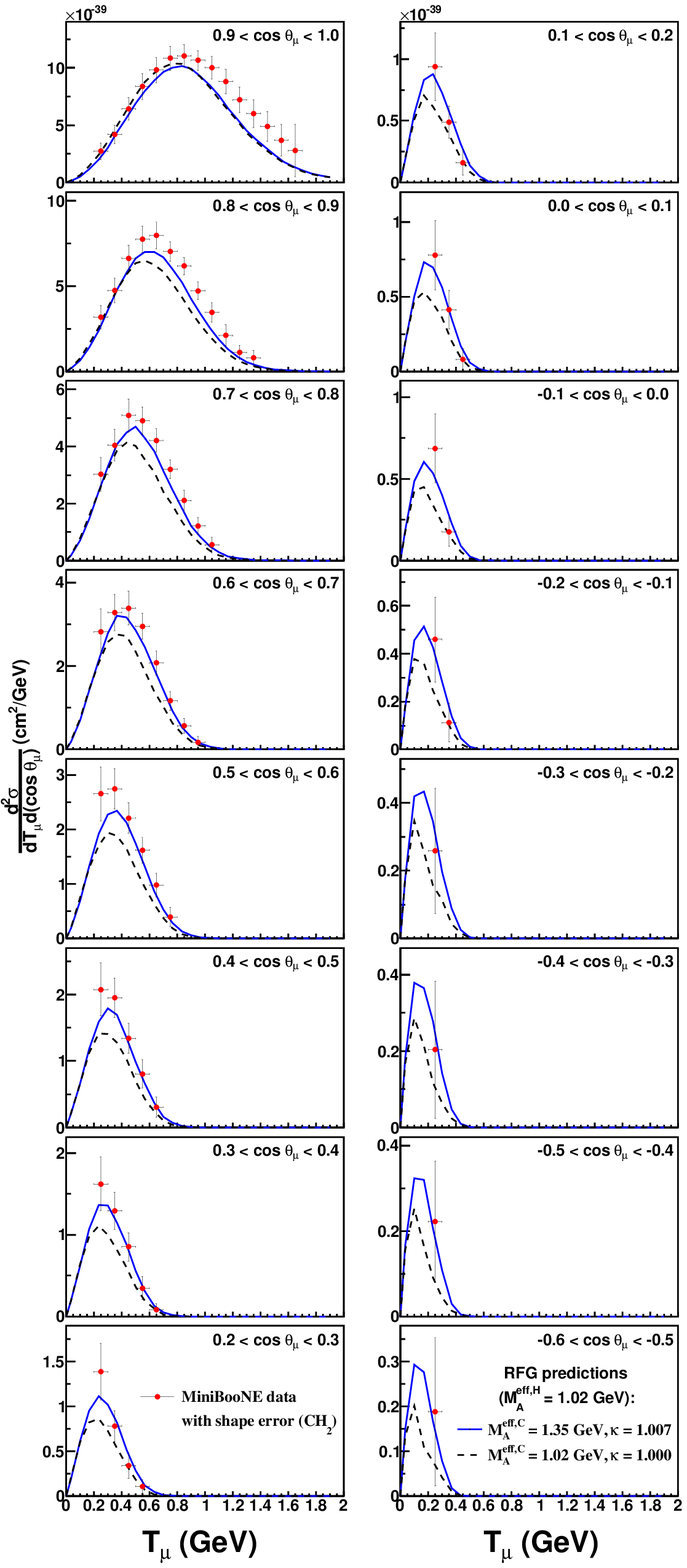} \\
\caption{(Color online) Projections of the per-proton double-differential cross section in muon kinetic energy $\tmu$ for various scattering angles $\uz$.  As indicated, both RFG predictions assume an effective axial mass of 1.02~GeV for the quasifree hydrogen component of the data, while two choices for CCQE model parameters are shown.  Data with $\uz <$ -0.6 have insufficient statistics and this region is not shown.  Shape uncertainties are shown; an additional normalization uncertainty of 13.0\% is not.  Numerical values for this cross section and its uncertainty are provided in Tables~\ref{tbl:dblDifflCH2} and~\ref{tbl:dblDifflShapeCH2}, respectively.}
\label{fig:twoDwComps}
\end{figure}

\section{\label{sec:conc} Conclusion}

This work presents the first measurement of the $\numub$ CCQE double-differential cross section in terms of muon angle and energy.  This measurement is also the first $\numub$ charged-current cross-section measurement with the majority of interactions with $E_{\bar{\nu}} < $ 1~GeV.  This cross section is the least model-dependent measurement possible with the MiniBooNE detector and is thus the main result of this work.  

It is clear in Figure~\ref{fig:twoDwComps} that the RFG model (described in Section~\ref{sec:nuInt}) assuming $\maeffc \sim 1$~GeV does not adequately describe these data in shape or in normalization.  Consistent with other recent CCQE measurements on nuclear material~\cite{qePRD,K2KsciFi,MINOS,SB}, a significant enhancement in the normalization that grows with decreasing muon scattering angle is observed compared to the expectation with $M_A$~=~1.0~GeV.

These data find tension with the NOMAD $\numub$ CCQE results, which are described both in shape and normalization by $M_A$ = 1.06 $\pm$ 0.12~\cite{NOMAD}.  This tension is common among the $\numu$ CCQE analyses from the two experiments.  However, care should be taken in comparing model-dependent results among experiments with such different neutrino fluxes and detector technologies.  A definitive unification of these apparently discrepant data sets will require the continued increase of both experimental and theoretical activity surrounding this topic.  Fortunately, many experiments at a variety of neutrino energies capable of making high-resolution, model-independent neutrino and antineutrino CCQE measurements with different detector technologies and nuclear media using both neutrino and antineutrino beams currently have data or will soon.  These include MINER$\nu$A~\cite{minerva}, SciBooNE~\cite{sbLoi}, MicroBooNE~\cite{uB}, ArgoNeuT~\cite{argoneut}, ICARUS~\cite{icarus} and the T2K~\cite{t2k} and NO$\nu$A~\cite{nova} near detectors. 

Finally, a novel and crucial evaluation of the $\numu$ background in this work is presented in Appendix~\ref{apndx:mumCap}.  In the absence of a magnetic field, this analysis and those described in Ref.~\cite{wsPRD} measure the $\numu$ flux of the antineutrino-mode beam with $\sim$ 15\% fractional uncertainty.  These techniques could be used in current and future neutrino oscillation programs, particularly when modest charge identification is sufficient to meet the physics goals~\cite{hubSch}.

This work was made possible with the support of Fermilab, the National Science Foundation, and the Department of Energy through the construction, operation, and data analysis of the MiniBooNE experiment.  Operated by Fermi Research Alliance, LLC under Contract No. De-AC02-07CH11359 with the United States Department of Energy.


\newpage
\clearpage

\appendix
\section{\label{apndx:mumCap} Measurement of $\numu$ flux in antineutrino mode using $\mum$ capture}

\subsection{\label{apndx:sbsec:intro} Introduction}

MiniBooNE uses dedicated hadroproduction measurements from the HARP experiment~\cite{HARP} to predict the $\numu$ and $\numub$ fluxes for the antineutrino-mode beam.  However,
as shown in Figure~\ref{fig:thPi} in Section~\ref{sbsec:mbFlux}, most of the $\numu$ flux arises from the very forward-going region of $\pip$ production and is not well constrained by the HARP measurements.  Fortunately, there are several ways to determine the $\numu$ content of the beam directly from MiniBooNE data.  Two such analyses are described in Ref.~\cite{wsPRD} and a third is presented in this appendix.
These analyses show that, in the absence of a magnetic field, the $\numu$ and $\numub$ content can still be modestly separated using statistical methods.


The measurement of the $\numu$ flux in the antineutrino-mode beam described in this appendix exploits the asymmetry in the production of decay electrons between $\mum$ and $\mup$ in nuclear material. The results are consistent with and complementary to those of Ref.~\cite{wsPRD}. 

\subsection{\label{apndx:sbsec:mumCap} Muon capture model and event selection}

The model for $\mum$ capture and the processes that can obscure its rate in the MiniBooNE detector is described in this section, followed by details on the analysis samples studied.  In mineral oil, stopped $\mum$ are captured on carbon nuclei with a probability of (7.78~$\pm$~0.07)\%~\cite{mumCap}. In such capture events, typically little or no extra activity is observed in the detector.  However, the low-energy neutron and photons from the primary capture reaction as well as deexcitations of the boron isotope may be energetic enough to produce a Michel-like event.  The simulated production of these particles is based on the measurements of Refs.~\cite{mumCap1,mumCap2,mumCap3,mumCap4,mumCap5,mumCap6}, and the model that propagates these particles and possible reinteractions through the MiniBooNE detector estimates 6.60\% of $\mum$ capture events lead to activity similar to a  low-energy Michel.  Thus, the apparent $\mum$ nuclear capture probability in the detector is predicted to be 7.78 $\times$ (100\% - 6.60\%) = 7.26 $\pm$ 0.20\%, where the uncertainty is substantially increased to recognize the model dependence of the rate to regain Michel-like events following $\mum$ capture.  This rate is partially constrained by the calibration procedure described in Section~\ref{apndx:sbsec:calib}, and it will be shown that the assigned uncertainty on effective $\mum$ nuclear capture has a negligible impact on the final measurements.

Sensitivity to the $\mum$ content of the data is obtained by simultaneously analyzing two samples: those with only a muon candidate event, and events consistent with a muon and its decay electron.  Therefore, this analysis takes as signal all $\numu$ and $\numub$ charged-current events.  Apart from the requirement of either one or two subevents, the event selection for this analysis closely follows that described in Section~\ref{sbsec:sel} with a few changes appropriate to different backgrounds and a higher sensitivity to Michel detection efficiency.  Table~\ref{tbl:mumCapSel} details the $\numu$ and $\numub$ charged-current purity of the two samples after each cut.

\begin{table*}
\caption{\label{tbl:mumCapSel} Antineutrino-mode purity in \% for all $\numu$ and $\numub$ charged-current events in the one- and two-subevent samples. A precut of generated radius $<$ 550~cm is applied.}
\begin{ruledtabular}
\begin{tabular}{cccccc}
\multirow{2}{*}{Cut \#} & \multirow{2}{*}{Description} & \multicolumn{2}{c}{One subevent} & \multicolumn{2}{c}{Two subevents}  \\
 & & $\numu$ CC & $\numub$ CC & $\numu$ CC & $\numub$ CC \\
\hline
1 & Subevent cut                            		 & 18 & 33 & 26 & 57 \\
2 & Veto hits \textless\,6 for all subevents 		 & 9 & 11 & 30 & 65 \\
3 & First subevent in beam window: $4000<T(\mathrm{ns})<7000$ & 9 & 11 & 29 & 65 \\
4 & Reconstructed vertex radius $<500$~cm for first subevent         & 8 & 11 & 29 & 65 \\
5 & Kinetic energy $>200$~MeV for first subevent under $\mu$ hypothesis   & 20 & 27 & 29 & 68 \\
6 & $\mu/e$ log-likelihood ratio $>0.02$ for first subevent           & 36 & 54 & 27 & 72 \\
7 & Predicted $\mu$ stopping radius $<$ 500~cm & 39 & 46 & 28 & 71 \\
8 & $\qsqqe$ $>$ 0.2 GeV$^{2}$ & 57 & 36 & 43 & 56 \\
\end{tabular}
\end{ruledtabular}
\end{table*}

\begin{figure}[h]
\includegraphics[scale=0.46]{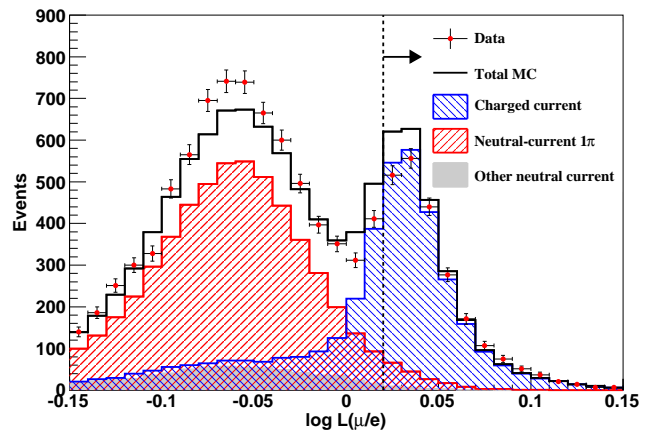} \\
\caption{(Color online) The log-likelihood $\mu/e$ particle-ID variable in the single-subevent sample.  Events with a muon-like score of 0.02 and higher are selected.  Expectations are normalized to flux, and errors shown on data are statistical only.}
\label{fig:1seLkCut}
\end{figure}

The primary samples of this analysis are separated by Cut 1, where $\numu$ CC events have an enhanced contribution in the single-subevent sample due to $\mum$ capture.  Cuts 2-5 are common to the analysis presented in the main body of this work and are motivated in Section~\ref{sbsec:sel}.  Cuts 6 and 8 reduce the neutral-current background in the single-subevent sample: Figure~\ref{fig:1seLkCut} shows neutral-current single $\pi$ events are largely rejected by the requirement on the $\mu/e$ log-likelihood variable, while cut 8 further reduces their contribution.  Cut 7 uses the observed muon kinematics and the stopping power of mineral oil for minimum-ionizing particles to calculate where the muon will stop.  This cut removes Michels produced near the optical barrier where Michel detection efficiency decreases rapidly with radius and is thus sensitive to modeling, while Michel detection is constant below 500~cm in this variable.  Cut 7 also enhances $\numu$ purity due to kinematic differences between $\numu$ and $\numub$ CCQE, where the more forward-going nature of the $\mup$ from $\numub$ interactions preferentially stop at high radius in the downstream region of the detector.  A summary of nucleon-level interactions contributing to the selected subevent samples is given in Table~\ref{tbl:mumCapPur}.

\begin{table}
\caption{\label{tbl:mumCapPur} Summary of predicted nucleon-level interactions in the antineutrino-mode subevent samples.  The small contribution from neutral-current processes are presented as the sum of the $\numu$ and $\numub$ interactions.}
\begin{ruledtabular}
\begin{tabular}{ccc}
\multirow{2}{*}{Process} & \multicolumn{2}{c}{Contribution (\%) to} \\
 & One subevent & Two subevents\\
\hline
$\numub p \to \mup n$ 								& 31 & 49 \\
$\numu n \to \mum p$ 									& 48 & 36 \\
$\numub N \to \mup N \pim$ 							& 3 & 5 \\
$\numu N \to \mum N \pip$						 		& 7 & 7 \\
$\numu (\numub) N \to \numu (\numub) N$ 				& 1 & 0 \\
$\numu (\numub) N \to \numu (\numub) N \piz$ 			& 3 & 0 \\
$\numu (\numub) N \to \numu (\numub) N \pipm$ 			& 4 & 0 \\
other 												& 3 & 3 \\
\end{tabular}
\end{ruledtabular}
\end{table}

\subsection{\label{apndx:sbsec:calib} Calibrations using neutrino-mode data}

Charged-current $\numu$ and $\numub$ events without final-state $\pip$ typically have two subevents: one from the primary $\mu$ and another from its decay electron.  Two effects determine the majority of the migration rate of these events from the two-subevent to the one-subevent sample: $\mum$ nuclear capture and detection efficiency for Michel electrons. Since an appreciable number ($\sim$7\%) of charged-current events enter the single-subevent sample due to Michel detection inefficiencies, the measurement of the $\numu$ content of the antineutrino-mode data is sensitive to the accuracy of both the Michel detection efficiency and the effective $\mum$ capture rate.  The rate of nondetection is mostly due to Michel production too close in time with the parent muon to be separated by the subevent definition.  This effect can be seen at low values of the timing difference distribution between the two subevents shown in Figure~\ref{fig:timing}, Section~\ref{sbsec:sel}.

Fortunately, the neutrino-mode data offer an opportunity to calibrate the migration rate between the subevent samples for $\numu$ charged-current events.  Due to a convolution of flux and cross-section effects~\cite{wsPRD}, the neutrino-mode subevent samples are mostly due to charged-current $\numu$ interactions.  Table~\ref{tbl:nuModePur} shows the predicted neutrino species and interaction contributions to the neutrino-mode subevent samples. With a high-purity $\numu$ charged-current sample, the accuracy of Michel detection and effective $\mum$ capture in simulation can be tested.  For charged-current $\numu$ events without final-state $\pip$ ($\numu$ CC), the number of events in the neutrino-mode one-subevent (1SE$^{\nu}$) and two-subevent (2SE$^{\nu}$) samples are given by:

\begin{table}
\caption{\label{tbl:nuModePur} A brief description of the neutrino mode subevent samples for the same selection described in the previous section.}
\begin{ruledtabular}
\begin{tabular}{ccc}
\multirow{2}{*}{Process} & \multicolumn{2}{c}{Contribution (\%) to} \\
 & One subevent & Two subevents\\
\hline
All $\numu$ charged-current	& 95.4 & 99.0  \\
All $\numub$ 				& 0.4 &  0.7 \\
All neutral current 			& 4.3 &  0.3 \\
\end{tabular}
\end{ruledtabular}
\end{table}

\begin{center}
\begin{eqnarray}
\label{eqn:1seNu}
\textrm{1SE}^{\nu} &=&  \numu \textrm{CC} \times (\delta + \beta(1 - \delta)) + \textrm{N}_1^\nu\\
\label{eqn:2seNu}
\textrm{2SE}^\nu &=&  \numu \textrm{CC} \times (1 - \delta - \beta(1 - \delta)) + \textrm{N}_2^\nu \end{eqnarray}
\par\end{center} 

\noindent where N$_1^\nu$ (N$_2^\nu$) is the neutral-current contribution to the 1SE (2SE) sample, $\delta$ is the Michel detection inefficiency and $\beta$ is the effective $\mum$ capture rate described previously.  The rate for Michel nondetection can be solved in terms of the effective $\mum$ capture rate and the small neutral-current contribution:

\begin{center}
\begin{equation}
\label{eqn:calib1}
\delta = \frac{\frac{\textrm{1SE}^{\nu} - \textrm{N}_1^\nu}{\textrm{1SE}^\nu + \textrm{2SE}^\nu - (\textrm{N}_1^\nu + \textrm{N}_2^\nu)}-\beta}{1-\beta}\end{equation}
\par\end{center} 

\noindent Noting the symmetry in Equations~\ref{eqn:1seNu} and~\ref{eqn:2seNu} between $\delta$ and $\beta$, Equation~\ref{eqn:calib1} can also express the effective $\mum$ capture rate in terms of Michel detection with $\delta \leftrightarrow \beta$.  Table~\ref{tbl:calibs} gives values of $\delta$ and $\beta$ from simulation and data based on the observed or predicted event rates in the 1SE$^{\nu}$ and 2SE$^\nu$ samples.

\begin{table}
\caption{\label{tbl:calibs} Calibration summary for Michel detection inefficiency ($\delta$) and the rate of effective $\mum$ nuclear capture ($\beta$).  Note that both processes cannot be simultaneously constrained.}
\begin{ruledtabular}
\begin{tabular}{cccc}
Process & data & MC & data/MC \\
\hline
$\delta$ & 0.073 & 0.074 & 0.98 \\
$\beta$ & 0.071 & 0.073 & 0.98 \\
\end{tabular}
\end{ruledtabular}
\end{table}

As the $\numu$ charged-current migration rate to the single-subevent sample is due to a convolution of Michel detection and effective $\mum$ capture, the processes cannot be simultaneously calibrated with the neutrino-mode data - that is, for example, the calibration of $\delta$ assumes the MC valuation of $\beta$ is correct.  Future experiments may have the ability to separate the two processes by examining the low-energy region of the Michel spectrum, where the contribution from events following $\mum$ capture is enhanced.  As the calibration results shown in Table~\ref{tbl:calibs} are quite mild and within systematic uncertainties, this procedure gives confidence in the ability to unambiguously measure the $\numu$ content of the antineutrino-mode data using $\mum$ capture.  

The high-statistics neutrino-mode data also allow for a stability check of the ratio of samples one subevent / two subevents, and four sequential sample periods are consistent within one standard deviation.

\subsection{\label{apndx:sbsec:meas} Measurement and systematic errors}

The $\numu$ flux is measured by adjusting the MC prediction of the $\numu$ and $\numub$ content to match the data in regions of reconstructed energy for the subevent samples.  Following the conventions of Equations~\ref{eqn:1seNu} and~\ref{eqn:2seNu} and introducing $\numub$ CC for the $\numub$ charged-current content, the predicted $\numu$ and $\numub$ contributions to the subevent samples in antineutrino mode are defined as

\begin{center}
\begin{eqnarray}
\label{eqn:aux1}
\nu^{\textrm{1SE}}_{\textrm{MC}} &=& \numu \textrm{CC} \times (\delta + \beta(1 - \delta)) \\
\label{eqn:aux2}
\nu^{\textrm{2SE}}_{\textrm{MC}} &=& \numu \textrm{CC} \times (1 - \delta - \beta(1 - \delta)) \\
\label{eqn:aux3}
\bar{\nu}^{\textrm{1SE}}_{\textrm{MC}} &=& \numub \textrm{CC} \times \delta \\
\label{eqn:aux4}
\bar{\nu}^{\textrm{2SE}}_{\textrm{MC}} &=& \numub \textrm{CC} \times (1 - \delta) \end{eqnarray}
\par\end{center} 

Then the single- (``1SE$^{\bar{\nu}}$") and two-subevent (``2SE$^{\bar{\nu}}$") data samples in antineutrino mode are given by 

\begin{center}
\begin{eqnarray}
\label{eqn:1seNub}
\textrm{1SE}^{\bar{\nu}} &=& \anu \times \nu^{\textrm{1SE}}_{\textrm{MC}} + \anub \times \bar{\nu}^{\textrm{1SE}}_{\textrm{MC}} + \textrm{N}_1^{\bar{\nu}} \\
\label{eqn:2seNub}
\textrm{2SE}^{\bar{\nu}} &=&  \anu \times \nu^{\textrm{2SE}}_{\textrm{MC}} + \anub \times \bar{\nu}^{\textrm{2SE}}_{\textrm{MC}} + \textrm{N}_2^{\bar{\nu}} \end{eqnarray}
\par\end{center} 

\noindent where $\anu$ and $\anub$ are scale factors for the $\numu$ and $\numub$ charged-current content, respectively, to be measured in this analysis and the neutral-current content (N$_2^{\bar{\nu}}$ and N$_1^{\bar{\nu}}$) include contributions from both $\numu$ and $\numub$.  Equations~\ref{eqn:1seNub} and~\ref{eqn:2seNub} can be solved for $\anu$ and $\anub$:

\begin{center}
\begin{eqnarray}
\label{eqn:anu!}
\anu &=& \frac{(\textrm{1SE}^{\bar{\nu}} - \textrm{N}_1^{\bar{\nu}}) \bar{\nu}^{\textrm{2SE}}_{\textrm{MC}} - (\textrm{2SE}^{\bar{\nu}} - \textrm{N}_2^{\bar{\nu}}) \bar{\nu}^{\textrm{1SE}}_{\textrm{MC}} }{\bar{\nu}^{\textrm{2SE}}_{\textrm{MC}} \nu^{\textrm{1SE}}_{\textrm{MC}}-\bar{\nu}^{\textrm{1SE}}_{\textrm{MC}} \nu^{\textrm{2SE}}_{\textrm{MC}}} \\
\label{eqn:anub!}
\anub &=& \frac{(\textrm{1SE}^{\bar{\nu}} - \textrm{N}_1^{\bar{\nu}}) \nu^{\textrm{2SE}}_{\textrm{MC}} - (\textrm{2SE}^{\bar{\nu}} - \textrm{N}_2^{\bar{\nu}}) \nu^{\textrm{1SE}}_{\textrm{MC}} }{\nu^{\textrm{2SE}}_{\textrm{MC}} \bar{\nu}^{\textrm{1SE}}_{\textrm{MC}}-\nu^{\textrm{1SE}}_{\textrm{MC}} \bar{\nu}^{\textrm{2SE}}_{\textrm{MC}}} \end{eqnarray}
\par\end{center} 

To check the modeling of the $\numu$ flux spectrum, this measurement is performed in three regions of reconstructed energy $\enuqe$, defined as  

\begin{center}
\begin{equation}
\label{eqn:EnuQE}
\enuqe = \frac{2 \left(M_{p} - E_{B}\right) E_{\mu} - \left( E^{2}_{B} - 2 M_{p} E_{B} + m^{2}_{\mu} + \De M^{2} \right)}{2\left[ \left(M_{p} - E_{B} \right) - E_{\mu} + p_{\mu}\,\text{cos}\,\theta_{\mu}\right]}\end{equation}
\par\end{center}

\noindent where, $E_{B} = 30$~MeV is the binding energy, $m_{\mu}$ is the muon mass, $\De M^{2} = M_{p}^{2} - M_{n}^{2}$, where $M_{n}$ ($M_{p}$) is the neutron (proton) mass, $p_{\mu}$ is the muon momentum, and $\thetmu$ is the outgoing muon angle relative to the incoming neutrino beam.  This reconstruction assumes $\numub$ CCQE interactions with at-rest, independently acting nucleons.  Though this is a model-dependent valuation of the neutrino energy, complicated further by the significant non-$\numub$ CCQE content, separating the samples into exclusive regions of $\enuqe$ nevertheless affords statistical sensitivity to the accuracy of the simulated flux spectrum.  The three energy regions explored are $\enuqe~<~0.9$~GeV, $\enuqe~\geq$ 0.9~GeV, and an inclusive sample. The statistics of the single-subevent sample prohibit the analysis of more than two exclusive $\enuqe$ regions.  As described in the previous section, the calibration from the neutrino-mode data is ambiguous between Michel detection and the effective $\mum$ capture model.  As these effects change the expectations for $\bar{\nu}^{\textrm{1SE}}_{\textrm{MC}}, \bar{\nu}^{\textrm{2SE}}_{\textrm{MC}}, \nu^{\textrm{1SE}}_{\textrm{MC}}$ and $\nu^{\textrm{2SE}}_{\textrm{MC}}$ in different ways, the measurement of $\anu$ and $\anub$ is, in principle, sensitive to which rate is calibrated.  In the absence of a compelling reason to choose one over the other, the final evaluations for $\anu$ and $\anub$ are taken to be the average of the two calculations assuming each rate is calibrated.  A calibration uncertainty spanning the difference in the two measurements is added to the systematic errors discussed next.  The central values for $\anu$ and $\anub$ are presented in Table~\ref{tbl:anuAnubCV}.  

\begin{table}
\caption{\label{tbl:anuAnubCV} Central-value results for scale factors relative to MC expectation for the $\numu$ and $\numub$ charged-current content of the antineutrino-mode data.}
\begin{ruledtabular}
\begin{tabular}{ccccc}
\multirow{2}{*}{Parameter} & Calibrated & \multicolumn{3}{c}{$\enuqe$ range (GeV)} \\
& process & $<$ 0.9 & $\geq$ 0.9 & All \\
\hline
\multirow{3}{*}{$\anu$} & $\delta$ & 0.78 & 0.79 & 0.78 \\
					   & $\beta$  & 0.78 & 0.79 & 0.78 \\
					   & {\bf Average}  & {\bf 0.78} & {\bf 0.79} & {\bf 0.78} \\
\hline
\multirow{3}{*}{$\anub$} & $\delta$ & 1.16 & 1.15 & 1.16 \\
					    & $\beta$  & 1.16 & 1.15 & 1.16 \\
					    & {\bf Average}  & {\bf 1.16} & {\bf 1.15} & {\bf 1.16} \\
\end{tabular}
\end{ruledtabular}
\end{table}

Systematic uncertainties on $\anu$ and $\anub$ are evaluated by assigning relevant errors to the physics processes contributing to the subevent samples and observing how the measurement changes as the channels are varied within their uncertainty.  These uncertainties are treated as uncorrelated, so the uncertainty on $\anu$, for example, due to physics processes $P_1, \dotsb, P_N$ is simply 

\begin{center}
\begin{eqnarray}
\label{eqn:systErrs}
\delta \anu^2 = \sum\limits_{i=1}^N \left( \frac{\partial \anu}{\partial P_i} \delta P_i \right)^2  \end{eqnarray}
\par\end{center} 

Table~\ref{tbl:muCapSysts} shows the errors assigned to the various contributing processes and their propagated uncertainty onto $\anu$ and $\anub$.  The most important process for extracting the $\numu$ flux measurement is the $\numu$ CCQE interaction, and its cross section and assigned uncertainty reflect the measurement and accuracy of the MiniBooNE result~\cite{qePRD}.  The same is true for the $\numu$ and $\numub$ neutral-current single $\piz$ channels~\cite{ncPizPRD}; however the error is increased to recognize a possible rate difference in these interactions between the cross-section measurements and this analysis due to using the opposite side of the log-likelihood variable shown in Figure~\ref{fig:1seLkCut}. The $\numu$ and $\numub$ charged-current single charged $\pi$ channels are adjusted to reflect the $\numu$ measurement~\cite{qePRD} and their uncertainty is increased to recognize the extrapolation to the $\numub$ processes.  Treating the uncertainties on the $\numu$ processes constrained by MiniBooNE data as uncorrelated ignores a common dependence on the neutrino-mode flux uncertainties and a small cancellation of errors that could be  propagated onto $\anu$ and $\anub$ is ignored.  The $\numu$ neutral-current elastic process is also constrained by MiniBooNE data~\cite{ncePRD}, while the neutral-current charged-pion production processes are completely unconstrained and so the assigned uncertainty is large.  Preliminary results for the $\numub$ CCQE process~\cite{JG_nuInt11} informs the choice of a 20\% uncertainty relative to the RFG model with $M_A$ = 1.35 GeV. With these systematic uncertainty assumptions, as seen in Table~\ref{tbl:muCapSysts}, the uncertainty on the main result of this work $\anu$ is dominated by statistics and the $\numu$ CCQE cross section.  As the $\numu$ CCQE process is directly constrained by MiniBooNE data, the measurement of the $\numu$ flux scale $\anu$ features negligible model dependence.  Table~\ref{tbl:finalMeas} summarizes the measurements of $\anu$ and $\anub$.

\begin{table*}
\caption{\label{tbl:muCapSysts} Uncertainty summary for this analysis.  Included are the assumed errors on physics processes and their contributions to the total errors in $\anu$ and $\anub$ in the regions of reconstructed neutrino energy studied.  The statistics of the $\nu$-mode data enter the uncertainty from the calibration procedure described in Section~\ref{sbsec:sel}.}
\begin{ruledtabular}
\begin{tabular}{cc|ccc|ccc}
\multirow{2}{*}{Source} & Fractional       & \multicolumn{3}{c|}{Uncertainty contribution to $\anu$}  & \multicolumn{3}{c}{Uncertainty contribution to $\anub$}  \\
& uncertainty (\%) & $\enuqe <$ 0.9 GeV & $\enuqe \geq$ 0.9 GeV & All & $\enuqe <$ 0.9 GeV & $\enuqe \geq$ 0.9 GeV & All \\
\hline
$\numu n \to \mum p$ 									& 10  & 0.07 & 0.08 & 0.07 & 0.00 & 0.00 & 0.00 \\
$\numub p \to \mup n$ 								& 20  & 0.04 & 0.02 & 0.03 & 0.20 & 0.20 & 0.21 \\
$\numu (\numub) N \to \mum (\mup) N \pip (\pim)$ 		& 20  & 0.04 & 0.05 & 0.04 & 0.02 & 0.02 & 0.01 \\
$\numu (\numub) N \to \numu (\numub) N$ 				& 30  & 0.00 & 0.00 & 0.00 & 0.00 & 0.00 & 0.00 \\
$\numu (\numub) N \to \numu (\numub) N \piz$ 			& 25  & 0.02 & 0.01 & 0.01 & 0.01 & 0.01 & 0.01 \\
$\numu (\numub) N \to \numu (\numub) N \pipm $ 		& 50  & 0.05 & 0.02 & 0.01 & 0.03 & 0.03 & 0.01 \\
$\mum$ capture 										& 2.8 & 0.00 & 0.00 & 0.00 & 0.00 & 0.00 & 0.00 \\
$\nubar$-mode statistics 								&  -  & 0.10 & 0.11 & 0.08 & 0.08 & 0.08 & 0.06 \\
$\nu$-mode statistics 								&  -  & 0.04 & 0.05 & 0.04 & 0.03 & 0.03 & 0.03 \\
\hline
{\bf{All}} 											& {\bf -} & {\bf 0.14} & {\bf 0.16} & {\bf 0.12} & {\bf 0.22} & {\bf 0.22} & {\bf 0.22} \\
\end{tabular}
\end{ruledtabular}
\end{table*}

\begin{table}
\caption{\label{tbl:finalMeas} Summary of measurements for the $\numu$ flux scale $\anu$ and the $\numub$ rate scale $\anub$.}
\begin{ruledtabular}
\begin{tabular}{cccc}
\multirow{2}{*}{Parameter} & \multicolumn{3}{c}{$\enuqe$ range (GeV)} \\
 & $<$ 0.9 & $\geq$ 0.9 & All \\
\hline
$\anu$ & 0.78 $\pm$ 0.14 & 0.79 $\pm$ 0.16 & 0.78 $\pm$ 0.12 \\
$\anub$ & 1.16 $\pm$ 0.22 & 1.15 $\pm$ 0.22 & 1.16 $\pm$ 0.22\\
\end{tabular}
\end{ruledtabular}
\end{table}

As the cross sections for the dominant $\numu$ processes have been applied to simulation, the deviation from unity for $\anu$ represents the accuracy of the highly uncertain $\numu$ flux prediction in antineutrino mode.  As the bulk of the $\numub$ flux prediction is constrained by the HARP data, the $\anub$ scale factor is representative of the level of cross-section agreement between the data and the RFG with $M_A$~=~1.35~GeV for the $\numub$ CCQE process.

\subsection{\label{apndx:sbsec:conc} Summary}

This appendix presents a measurement of the $\numu$ flux in antineutrino mode using a nonmagnetized detector.  The results are consistent with and complementary to the two measurements in Ref.~\cite{wsPRD}.  A summary of the results from all three analyses is shown in Figure~\ref{fig:wsSummary}, Section~\ref{sbsec:cons}.  As no energy dependence among the measurements is observed, the simulation of the $\numu$ flux in antineutrino mode, which is unconstrained by the HARP hadroproduction data, appears to be roughly 20\% high in normalization, while the flux spectrum is well modeled.

These techniques could also aid future neutrino experiments that will test for CP violation in the lepton sector using large unmagnetized detectors such as NO$\nu$A~\cite{nova}, T2K~\cite{t2k}, LBNE~\cite{lbne}, LAGUNA~\cite{memphys}, and Hyper-K~\cite{hyperk}.  In particular, the precision of $\sim$ 15\% in the determination of the $\numu$ flux of the antineutrino-mode beam using $\mum$ capture obtained here could easily be surpassed and the flux spectrum more rigorously checked by future experiments housing heavier nuclei.  As an example, the $\mum$ capture rate on $^{40}$Ar exceeds 70\%~\cite{mumCap}, almost affording event-by-event discrimination of the $\mu$ charge without a magnetic field.  Detector-specific complications arising from $\pi$/$\mu$ identification and Michel detection should not reduce sensitivity to the $\mu$ charge dramatically.

\section{\label{apndx:modDepCH2} Model-dependent measurements for $\numub$ interactions on CH$_2$}

This appendix presents MiniBooNE $\numub$ CCQE cross-section measurements that are explicitly dependent on CCQE interaction assumptions.  These measurements include all $\numub$ CCQE interactions as signal, while Appendix~\ref{apndx:modDepC} gives cross sections treating the hydrogen CCQE component as background.  All results are tabulated in Appendix~\ref{apndx:tables}.

\subsection{\label{apndx:modDepCH2:enu} Total cross section}

As the energy distribution of the incident $\numub$ beam is quite broad (Figure~\ref{fig:nubFlux}), the {\it a priori} knowledge of the neutrino energy is highly uncertain on an event-by-event basis.  If hadronic reconstruction is unavailable, it is typical for neutrino experiments to reconstruct the neutrino energy of events in the CCQE sample assuming scattering off of at-rest and independently-acting nucleons (``$\enuqe$") based solely on the outgoing lepton kinematics (Equation~\ref{eqn:EnuQE}).  Finding the neutrino energy in this way is often used to measure neutrino oscillation parameters, in particular the mass splitting, and it has been argued elsewhere that the assumptions implicit in this reconstruction significantly bias these measurements due to ignored nuclear effects~\cite{martiniOsc,moselEnuReco}.

Apart from the bias in the reconstructed energy distribution, a measurement of the absolute cross section over the observed energy range additionally suffers from model dependence through the unfolding procedure.  The total cross section is typically computed by unfolding the reconstructed neutrino energy to the ``true" energy distribution, and this correction is dependent on both the nuclear model used and detector resolution effects.  This is the main reason MiniBooNE has generally opted to report cross sections in terms of observed kinematics.  Due to these measurement biases, the MiniBooNE $\numub$ CCQE absolute cross section is not the main result of the work but is provided here for historical comparisons.  

A consequence of the unfolding bias is that one should exercise caution in comparing theoretical calculations to these results.  A strict comparison with these data and an external model involves finding the total cross section as a function of $\enuqe$ using the generated muon kinematics, and subsequently unfolding this distribution according to the RFG.  An example of this procedure can be found in Ref.~\cite{nievesUnfold}.

\begin{figure}[h]
\includegraphics[scale=0.46]{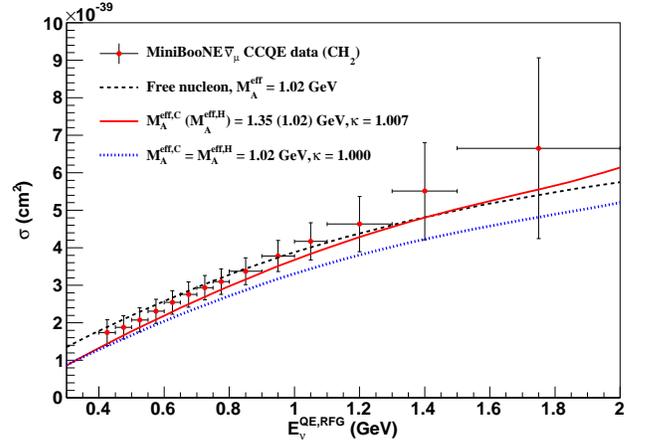} \\
\caption{(Color online) Per-nucleon total cross section for MiniBooNE $\numub$ CCQE data including the hydrogen scattering component.  The distribution is labeled $E_{\nub}^{QE,RFG}$ to recognize the dependence on the assumptions inherent in both the reconstruction and in the unfolding model.  Total errors are shown with data.  Numerical values are provided in Table~\ref{tbl:absXsecCH2}.}
\label{fig:absXsecCH2}
\end{figure}

The flux-unfolded $\numub$ CCQE cross section per nucleon is calculated assuming:

\begin{center}
\begin{equation}
\label{eqn:absXsec}
\sigma_i = \frac{\sum_j U_{ij}\left(d_j - b_j\right)}{\epsilon_i \,\Phi_i \,N}, \end{equation}
\par\end{center} 

\noindent where the same conventions used in Equation~\ref{eqn:dblDiffl} apply here with a few exceptions: as mentioned, the unfolding matrix $U_{ij}$ here connects the reconstructed neutrino energy (inferred from the observed $\mu$ kinematics via Equation~\ref{eqn:EnuQE}) to the generated distribution, and the flux term $\Phi_i$ refers to the $\numub$ flux exclusive to the $i^{\textrm{th}}$ neutrino energy bin.  Figure~\ref{fig:absXsecCH2} compares the observed total cross section to a few predictions from the RFG.  

\subsection{\label{apndx:modDepCH2:qsq} Momentum transfer}

Another important quantity for CCQE interactions is the squared four-momentum transfer $Q^{2} = \left(p_\nu - p_\mu \right)^2$.  However, again ignorance of the incoming neutrino energy prevents a clean measurement of this variable.  As in the case for $\enuqe$ (Equation~\ref{eqn:EnuQE}), if only lepton kinematics are available the distribution can be inferred by assuming CCQE scattering with an at-rest, independent nucleon:

\begin{center}
\begin{equation}
\label{eqn:qsqqe}
Q^{2}_{QE} = -m_\mu^2 + 2 \enuqe \left( E_\mu - p_\mu \uz \right)\end{equation}
\par\end{center} 

\noindent where $E_\mu$, $p_\mu$ and $m_\mu$ refer to the muon energy, momentum and mass, respectively.  The value of the axial mass is typically extracted from the shape of this distribution, so the differential cross section with respect to this variable is provided for historical comparisons despite the reconstruction assumptions.  However, to minimize the model dependence of this cross-section configuration, the reconstructed distribution of $\qsqqe$ is corrected to \emph{true} $\qsqqe$ - that is, Equation~\ref{eqn:qsqqe} with the generated muon kinematics.  In this way, the unfolding procedure only corrects for muon resolution effects and is not biased by the CCQE interaction model.  Note that truth-level $\qsqqe$ is only the same as the squared four-momentum transfer up to the naive reconstruction assumptions.  This choice is not typically made and so comparisons with similar cross sections from other experiments should be made with care.

\begin{figure}[h]
\includegraphics[scale=0.46]{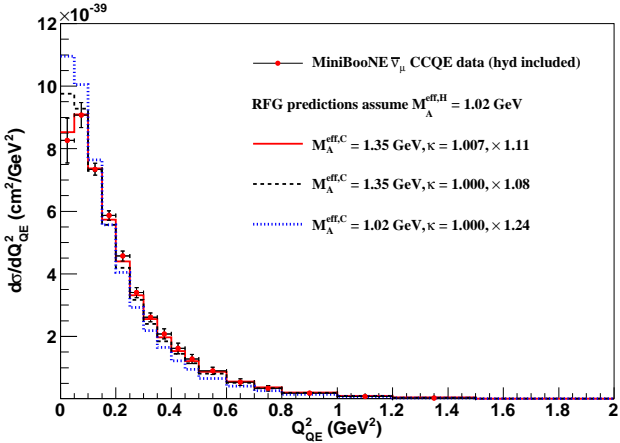} \\
\caption{(Color online) Per-nucleon single-differential cross section for MiniBooNE $\numub$ CCQE data including hydrogen CCQE events.  The RFG predictions are normalized to the observed total cross section $\int \frac{d\sigma}{d\qsqqe}d\qsqqe$ and the relative scales are indicated.  All predictions assume an effective axial mass of 1.026~GeV for the hydrogen scattering component.  Shape errors are shown with data.  Numerical values are provided in Table~\ref{tbl:qsqqeCH2}.}
\label{fig:qsqqeCh2}
\end{figure}

The flux-folded, single-differential cross section $d\sigma/d\qsqqe$ calculated in the same manner as the double-differential cross section (Equation~\ref{eqn:dblDiffl}) but for a single dimension:

\begin{center}
\begin{equation}
\label{eqn:qsqqeSigma}
\left(\frac{d\sigma}{d\qsqqe}\right)_i = \frac{\sum_j U_{ij}\left(d_j - b_j\right)}{(\Delta \qsqqe)_i\,\epsilon_i \,\Phi \,N}, \end{equation}
\par\end{center} 

\noindent where the same conventions used in Equation~\ref{eqn:dblDiffl} apply.  Figure~\ref{fig:qsqqeCh2} compares the results with shape uncertainty to predictions from the RFG normalized to data.

The conventions used to calculate $\sigma$(E$_{\nu}$) and $\frac{d\sigma}{d\qsqqe}$ are the same used to calculate the corresponding $\numu$ CCQE cross sections reported in Ref.~\cite{qePRD}.

\section{\label{apndx:modDepC} Model-dependent measurements for $\numub$ CCQE interactions on carbon}

\begin{figure}[h]
\includegraphics[scale=0.46]{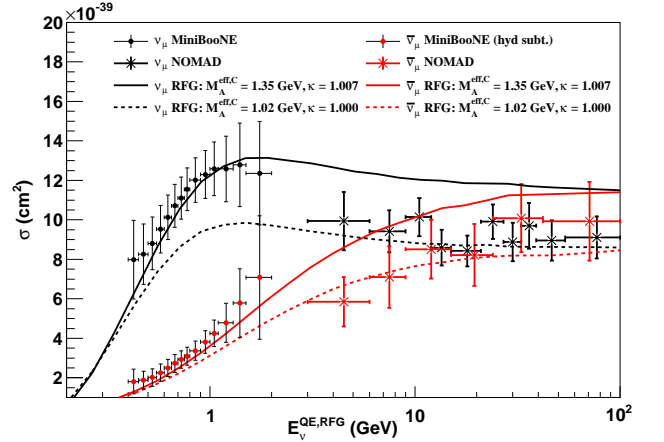} \\
\caption{(Color online) Total $\numu$ and $\numub$ CCQE cross sections for the MiniBooNE and NOMAD experiments, shown with two choices for the axial mass in the RFG for comparison.  The hydrogen content has been subtracted from the MiniBooNE $\numub$ data, and total uncertainties are shown.  Note the two experiments use difference detector technologies and so naturally assume different topologies in defining CCQE events.  Therefore, conclusions should be drawn with care.  NOMAD data taken from Ref.~\cite{NOMAD}, and MiniBooNE $\numu$ data taken from Ref.~\cite{qePRD}.  Numerical values for the MiniBooNE $\numub$ cross section are provided in Table~\ref{tbl:absXsec12C}.}
\label{fig:mbNomad}
\end{figure}

Following the same definitions for the total and single-differential (Equations~\ref{eqn:absXsec}, and~\ref{eqn:qsqqeSigma}, respectively) cross sections, results for $\numub$ CCQE on carbon are obtained following the subtraction of $\numub$ CCQE events on quasifree protons assuming $M_{A}~=~1.026$ GeV.  In this configuration, the total per-nucleon cross section for CCQE interactions on carbon from both the MiniBooNE $\numu$ and $\numub$ analyses may be compared to the corresponding NOMAD results, and this is shown in Figure~\ref{fig:mbNomad}.  

\section{\label{apndx:tables} Tabulation of results}

This appendix provides numerical values for the observed $\numub$ flux and all cross-section results presented in this work.  In addition, each cross section is accompanied by both the $\numub$ ``CCQE-like" and the CC1$\pim$ backgrounds subtracted from the data in the procedure to obtain the $\numub$ CCQE cross sections.  The CC1$\pim$ background is a subset of the $\numub$ CCQE-like background and is dominant in most regions.  Note that in order to facilitate comparisons with the predictions of $\numub$ CCQE and CCQE-like processes, the CCQE-like measurements exclude the $\numu$ content of the subtracted data.  The cross sections for these background processes are calculated for the various cross sections (Equations~\ref{eqn:dblDiffl},~\ref{eqn:absXsec}, and~\ref{eqn:qsqqe}) by replacing ($d_j - b_j$) with the appropriate subset of $b_j$: in the case of the CC1$\pim$, included are all resonance and coherent CC1$\pim$ as predicted by the Rein-Sehgal model~\cite{R-S}, while the CCQE-like cross sections include all background $\numub$ processes.  Note also these measurements are normalized to the total number of proton targets in the detector, even though the dominant interaction of CC1$\pim$ has nucleon-level interactions with neutrons as well.  This configuration is chosen for consistency with the $\numu$ CCQE-like background measurements, which were normalized to the number of neutron targets in the $\numu$ CCQE analysis~\cite{qePRD}.  As the CCQE-like cross sections on mineral oil and carbon differ only by the inclusion of the hydrogen content, the amount of $\numub$ hydrogen CCQE subtracted from the data (in the case of the latter calculation) can be found by taking the difference of these two cross sections.  To find the calculated per-nucleon $\numub$ hydrogen CCQE cross section, this difference should also be scaled by the ratio of total protons targets to quasifree proton targets, $2.03 \times 10^{32} / 0.70 \times 10^{32}$ = 2.9.

\subsection{\label{apndx:sbsec:flux} Antineutrino mode fluxes}

Section~\ref{sbsec:mbFlux} describes the flux prediction, and Table~\ref{tbl:nubflux} (Table~\ref{tbl:nuflux}) lists the predicted $\numub$ ($\numu$) flux in antineutrino-mode running per POT in 50~MeV wide bins of energy up to 3~GeV.  These values normalized to the observed exposure of 10.1 $\times$ 10$^{20}$ POT are shown in Figure~\ref{fig:nubFlux}.

\begin{table*}
\caption{
Predicted $\numub$ flux at the MiniBooNE detector in antineutrino mode.}
\begin{tabular}{cccc|cccc|cccc}
\hline
\hline
&$E_\nu$ bin && $\numub$ flux           &&$E_\nu$ bin && $\numub$ flux           &&$E_\nu$ bin && $\numub$ flux \\
&(GeV)       && ($\numub$/POT/50 MeV/cm$^2$) &&(GeV)       && ($\numub$/POT/50 MeV/cm$^2$) &&(GeV)       && ($\numub$/POT/50 MeV/cm$^2$) \\
\hline
&0.00-0.05&&$2.157\times 10^{-12}$&&1.00-1.05&&$7.658\times 10^{-12}$&&2.00-2.05&&$2.577\times 10^{-13}$\\
&0.05-0.10&&$7.840\times 10^{-12}$&&1.05-1.10&&$6.907\times 10^{-12}$&&2.05-2.10&&$2.066\times 10^{-13}$\\ 
&0.10-0.15&&$9.731\times 10^{-12}$&&1.10-1.15&&$6.180\times 10^{-12}$&&2.10-2.15&&$1.665\times 10^{-13}$\\ 
&0.15-0.20&&$1.141\times 10^{-11}$&&1.15-1.20&&$5.505\times 10^{-12}$&&2.15-2.20&&$1.346\times 10^{-13}$\\ 
&0.20-0.25&&$1.319\times 10^{-11}$&&1.20-1.25&&$4.877\times 10^{-12}$&&2.20-2.25&&$1.081\times 10^{-13}$\\ 
&0.25-0.30&&$1.438\times 10^{-11}$&&1.25-1.30&&$4.269\times 10^{-12}$&&2.25-2.30&&$8.837\times 10^{-14}$\\ 
&0.30-0.35&&$1.477\times 10^{-11}$&&1.30-1.35&&$3.686\times 10^{-12}$&&2.30-2.35&&$7.136\times 10^{-14}$\\ 
&0.35-0.40&&$1.479\times 10^{-11}$&&1.35-1.40&&$3.151\times 10^{-12}$&&2.35-2.40&&$5.707\times 10^{-14}$\\ 
&0.40-0.45&&$1.500\times 10^{-11}$&&1.40-1.45&&$2.678\times 10^{-12}$&&2.40-2.45&&$4.620\times 10^{-14}$\\ 
&0.45-0.50&&$1.485\times 10^{-11}$&&1.45-1.50&&$2.262\times 10^{-12}$&&2.45-2.50&&$3.778\times 10^{-14}$\\ 
&0.50-0.55&&$1.447\times 10^{-11}$&&1.50-1.55&&$1.898\times 10^{-12}$&&2.50-2.55&&$3.028\times 10^{-14}$\\ 
&0.55-0.60&&$1.406\times 10^{-11}$&&1.55-1.60&&$1.580\times 10^{-12}$&&2.55-2.60&&$2.412\times 10^{-14}$\\ 
&0.60-0.65&&$1.345\times 10^{-11}$&&1.60-1.65&&$1.311\times 10^{-12}$&&2.60-2.65&&$1.977\times 10^{-14}$\\ 
&0.65-0.70&&$1.287\times 10^{-11}$&&1.65-1.70&&$1.083\times 10^{-12}$&&2.65-2.70&&$1.638\times 10^{-14}$\\ 
&0.70-0.75&&$1.221\times 10^{-11}$&&1.70-1.75&&$8.917\times 10^{-13}$&&2.70-2.75&&$1.323\times 10^{-14}$\\ 
&0.75-0.80&&$1.152\times 10^{-11}$&&1.75-1.80&&$7.285\times 10^{-13}$&&2.75-2.80&&$1.038\times 10^{-14}$\\ 
&0.80-0.85&&$1.075\times 10^{-11}$&&1.80-1.85&&$5.941\times 10^{-13}$&&2.80-2.85&&$8.707\times 10^{-15}$\\ 
&0.85-0.90&&$9.980\times 10^{-12}$&&1.85-1.90&&$4.834\times 10^{-13}$&&2.85-2.90&&$6.981\times 10^{-15}$\\ 
&0.90-0.95&&$9.177\times 10^{-12}$&&1.90-1.95&&$3.937\times 10^{-13}$&&2.90-2.95&&$6.078\times 10^{-15}$\\ 
&0.95-1.00&&$8.411\times 10^{-12}$&&1.95-2.00&&$3.180\times 10^{-13}$&&2.95-3.00&&$5.111\times 10^{-15}$\\ 
\hline
\hline
\end{tabular}
\label{tbl:nubflux}
\end{table*}

\begin{table*}
\caption{
Predicted $\numu$ flux at the MiniBooNE detector in antineutrino mode.  Note that, based on the results of Ref.~\cite{wsPRD} and Appendix~\ref{apndx:mumCap}, the $\numu$ flux spectrum given here should be scaled by 0.77 to reflect the data-based constraints.}
\begin{tabular}{cccc|cccc|cccc}
\hline
\hline
&$E_\nu$ bin && $\numu$ flux           &&$E_\nu$ bin && $\numu$ flux           &&$E_\nu$ bin && $\numu$ flux \\
&(GeV)       && ($\numu$/POT/50 MeV/cm$^2$) &&(GeV)       && ($\numu$/POT/50 MeV/cm$^2$) &&(GeV)       && ($\numu$/POT/50 MeV/cm$^2$) \\
\hline

&0.00-0.05&&$2.298 \times 10^{-12}$&&1.00-1.05&&$1.087\times 10^{-12}$&&2.00-2.05&&$1.886\times 10^{-13}$\\ 
&0.05-0.10&&$5.903 \times 10^{-12}$&&1.05-1.10&&$1.044\times 10^{-12}$&&2.05-2.10&&$1.669\times 10^{-13}$\\ 
&0.10-0.15&&$3.726 \times 10^{-12}$&&1.10-1.15&&$9.967\times 10^{-13}$&&2.10-2.15&&$1.486\times 10^{-13}$\\ 
&0.15-0.20&&$2.338 \times 10^{-12}$&&1.15-1.20&&$9.435\times 10^{-13}$&&2.15-2.20&&$1.310\times 10^{-13}$\\ 
&0.20-0.25&&$2.570 \times 10^{-12}$&&1.20-1.25&&$8.826\times 10^{-13}$&&2.20-2.25&&$1.171\times 10^{-13}$\\ 
&0.25-0.30&&$1.797 \times 10^{-12}$&&1.25-1.30&&$8.320\times 10^{-13}$&&2.25-2.30&&$1.030\times 10^{-13}$\\ 
&0.30-0.35&&$1.776 \times 10^{-12}$&&1.30-1.35&&$7.736\times 10^{-13}$&&2.30-2.35&&$9.279\times 10^{-14}$\\ 
&0.35-0.40&&$1.855 \times 10^{-12}$&&1.35-1.40&&$7.180\times 10^{-13}$&&2.35-2.40&&$8.199\times 10^{-14}$\\ 
&0.40-0.45&&$1.834 \times 10^{-12}$&&1.40-1.45&&$6.609\times 10^{-13}$&&2.40-2.45&&$7.353\times 10^{-14}$\\ 
&0.45-0.50&&$1.770 \times 10^{-12}$&&1.45-1.50&&$6.053\times 10^{-13}$&&2.45-2.50&&$6.577\times 10^{-14}$\\ 
&0.50-0.55&&$1.701 \times 10^{-12}$&&1.50-1.55&&$5.533\times 10^{-13}$&&2.50-2.55&&$5.830\times 10^{-14}$\\ 
&0.55-0.60&&$1.618 \times 10^{-12}$&&1.55-1.60&&$5.058\times 10^{-13}$&&2.55-2.60&&$5.318\times 10^{-14}$\\ 
&0.60-0.65&&$1.555 \times 10^{-12}$&&1.60-1.65&&$4.577\times 10^{-13}$&&2.60-2.65&&$4.822\times 10^{-14}$\\ 
&0.65-0.70&&$1.493 \times 10^{-12}$&&1.65-1.70&&$4.134\times 10^{-13}$&&2.65-2.70&&$4.317\times 10^{-14}$\\ 
&0.70-0.75&&$1.425 \times 10^{-12}$&&1.70-1.75&&$3.725\times 10^{-13}$&&2.70-2.75&&$3.997\times 10^{-14}$\\ 
&0.75-0.80&&$1.357 \times 10^{-12}$&&1.75-1.80&&$3.336\times 10^{-13}$&&2.75-2.80&&$3.619\times 10^{-14}$\\ 
&0.80-0.85&&$1.302 \times 10^{-12}$&&1.80-1.85&&$3.003\times 10^{-13}$&&2.80-2.85&&$3.375\times 10^{-14}$\\ 
&0.85-0.90&&$1.236 \times 10^{-12}$&&1.85-1.90&&$2.663\times 10^{-13}$&&2.85-2.90&&$3.050\times 10^{-14}$\\ 
&0.90-0.95&&$1.192 \times 10^{-12}$&&1.90-1.95&&$2.375\times 10^{-13}$&&2.90-2.95&&$2.926\times 10^{-14}$\\ 
&0.95-1.00&&$1.141 \times 10^{-12}$&&1.95-2.00&&$2.126\times 10^{-13}$&&2.95-3.00&&$2.705\times 10^{-14}$ \\ 
\hline
\hline
\end{tabular}
\label{tbl:nuflux}
\end{table*}

\subsection{\label{apndx:sbsec:ch2} Cross-section results on CH$_2$}

All measurements in this section include the quasifree hydrogen CCQE scattering component and is therefore less model dependent compared to the results given in Appendix~\ref{apndx:sbsec:12c}, where the RFG model is relied on to subtract their contribution.  Shape uncertainties are provided for the double- and single-differential cross-section measurements, and these values should be used along with the total normalization uncertainty of 13.0\% in the context of a fit to these distributions.  The total uncertainty, including errors affecting both shape and normalization, is provided for the total cross section.


Numerical values for the MiniBooNE $\numub$ CCQE cross section including the hydrogen content is given in Table~\ref{tbl:dblDifflCH2}, while Table~\ref{tbl:dblDifflShapeCH2} provides the uncertainty on the shape of these data.  These tables correspond to Figures~\ref{fig:twoD} and~\ref{fig:twoDwComps}.  The CCQE-like and CC1$\pim$ backgrounds are reported in Tables~\ref{tbl:dblDifflQElikeCH2} and~\ref{tbl:dblDifflCCpim}, respectively.

\begin{table*}
\begin{center}
\caption{
The MiniBooNE $\numub$ CCQE double-differential cross section on mineral oil 
in units of $10^{-41}~\ucmt/\uGeV$.  Data is given in 0.1~GeV bins of $T_\mu$ (columns) and 
0.1~bins of $\uz$ (rows).}
\label{tbl:dblDifflCH2}
\scriptsize\addtolength{\tabcolsep}{-0.8pt}
\begin{tabular}{c|cccccccccccccccccc}
\hline
\hline
\raisebox{-0.5ex}{$\cos\th_\mu$}\raisebox{0.5ex}{$T_\mu$(GeV)}&0.2,0.3&0.3,0.4&0.4,0.5&0.5,0.6&0.6,0.7&0.7,0.8&0.8,0.9&0.9,1.0&
                                              1.0,1.1&1.1,1.2&1.2,1.3&1.3,1.4&1.4,1.5&1.5,1.6&1.6,1.7&1.7,1.8&1.8,1.9&1.9,2.0\\
\hline

+0.9,+1.0  	&  272.7  &  419.9  &  641.2  &  838.5  &  981.3  &  1083  &  1105  &  1065 &  1002  &  880.9  &  720.6  &  600.9  &  491.0  &  370.1  &  279.2  & ---  & ---  & --- \\ 
+0.8,+0.9  	&  319.4  &  474.1  &  662.8  &  773.1  &  795.7  &  702.5  &  616.6  &  471.3  &  346.2  &  211.1  &  111.7  &  79.70  & ---  & ---  & ---  & ---  & ---  & --- \\ 
+0.7,+0.8  	&  302.3  &  404.9  &  509.2  &  490.4  &  421.7  &  320.4  &  210.7  &  121.1  &  54.78  & ---  & ---  & ---  & ---  & ---  & ---  & ---  & ---  & --- \\ 
+0.6,+0.7  	&  281.4  &  328.1  &  338.7  &  295.0  &  207.3  &  116.4  &  56.66  &  17.61  & ---  & ---  & ---  & ---  & ---  & ---  & ---  & ---  & ---  & --- \\ 
+0.5,+0.6  	&  264.8  &  274.1  &  220.6  &  161.9  &  97.88  &  39.25  & ---  & ---  & ---  & ---  & ---  & ---  & ---  & ---  & ---  & ---  & ---  & --- \\ 
+0.4,+0.5  	&  207.5  &  195.0  &  133.9  &  80.43  &  30.57  & ---  & ---  & ---  & ---  & ---  & ---  & ---  & ---  & ---  & ---  & ---  & ---  & --- \\ 
+0.3,+0.4  	&  162.3  &  129.6  &  85.33  &  34.71  &  8.059  & ---  & ---  & ---  & ---  & ---  & ---  & ---  & ---  & ---  & ---  & ---  & ---  & --- \\ 
+0.2,+0.3 	&  138.7  &  78.16  &  33.78  &  10.84  & ---  & ---  & ---  & ---  & ---  & ---  & ---  & ---  & ---  & ---  & ---  & ---  & ---  & --- \\ 
+0.1,+0.2  	&  93.62  &  48.77  &  16.22  & ---  & ---  & ---  & ---  & ---  & ---  & ---  & ---  & ---  & ---  & ---  & ---  & ---  & ---  & --- \\ 
0.0,+0.1  		&  77.92  &  41.26  &  7.966  & ---  & ---  & ---  & ---  & ---  & ---  & ---  & ---  & ---  & ---  & ---  & ---  & ---  & ---  & --- \\ 
-0.1,0.0 		&  68.75  &  17.58  & ---  & ---  & ---  & ---  & ---  & ---  & ---  & ---  & ---  & ---  & ---  & ---  & ---  & ---  & ---  & --- \\ 
-0.2,-0.1 	&  45.94  &  11.17  & ---  & ---  & ---  & ---  & ---  & ---  & ---  & ---  & ---  & ---  & ---  & ---  & ---  & ---  & ---  & --- \\ 
-0.3,-0.2 	&  25.86  & ---  & ---  & ---  & ---  & ---  & ---  & ---  & ---  & ---  & ---  & ---  & ---  & ---  & ---  & ---  & ---  & --- \\ 
-0.4,-0.3 	&  20.34  & ---  & ---  & ---  & ---  & ---  & ---  & ---  & ---  & ---  & ---  & ---  & ---  & ---  & ---  & ---  & ---  & --- \\ 
-0.5,-0.4 	&  22.19  & ---  & ---  & ---  & ---  & ---  & ---  & ---  & ---  & ---  & ---  & ---  & ---  & ---  & ---  & ---  & ---  & --- \\ 
-0.6,-0.5 	&  18.87  & ---  & ---  & ---  & ---  & ---  & ---  & ---  & ---  & ---  & ---  & ---  & ---  & ---  & ---  & ---  & ---  & --- \\ 
-0.7,-0.6 	& ---  & ---  & ---  & ---  & ---  & ---  & ---  & ---  & ---  & ---  & ---  & ---  & ---  & ---  & ---  & ---  & ---  & --- \\ 
-0.8,-0.7 	& ---  & ---  & ---  & ---  & ---  & ---  & ---  & ---  & ---  & ---  & ---  & ---  & ---  & ---  & ---  & ---  & ---  & --- \\ 
-0.9,-0.8 	& ---  & ---  & ---  & ---  & ---  & ---  & ---  & ---  & ---  & ---  & ---  & ---  & ---  & ---  & ---  & ---  & ---  & --- \\ 
-1.0,-0.9  	& ---  & ---  & ---  & ---  & ---  & ---  & ---  & ---  & ---  & ---  & ---  & ---  & ---  & ---  & ---  & ---  & ---  & --- \\ 
\hline
\hline
\end{tabular}
\end{center}
\end{table*}

\begin{table*}
\begin{center}
\caption{
Shape uncertainty in units of 10$^{-41}$~cm$^2$/GeV on the MiniBooNE $\numub$ CCQE double-differential cross section on mineral oil.  The total normalization uncertainty is 13.0\%.}
\label{tbl:dblDifflShapeCH2}
\scriptsize\addtolength{\tabcolsep}{-0.8pt}
\begin{tabular}{c|cccccccccccccccccc}
\hline
\hline
\raisebox{-0.5ex}{$\cos\th_\mu$}\raisebox{0.5ex}{$T_\mu$(GeV)}&0.2,0.3&0.3,0.4&0.4,0.5&0.5,0.6&0.6,0.7&0.7,0.8&0.8,0.9&0.9,1.0&
                                              1.0,1.1&1.1,1.2&1.2,1.3&1.3,1.4&1.4,1.5&1.5,1.6&1.6,1.7&1.7,1.8&1.8,1.9&1.9,2.0\\
\hline

+0.9,+1.0  &  74.16  &  80.76  &  98.56  &  112.5  &  109.4  &  104.4  &  95.49  &  86.86  &  98.66  &  108.3  &  111.8  &  121.4  &  130.7  &  139.2  &  226.3  & --- & --- & ---\\ 
+0.8,+0.9  &  68.92  &  73.05  &  78.25  &  75.73  &  78.11  &  57.11  &  50.74  &  51.42  &  55.36  &  63.12  &  41.93  &  44.40  & --- & --- & --- & --- & --- & ---\\ 
+0.7,+0.8  &  59.43  &  55.98  &  57.48  &  47.88  &  39.73  &  32.54  &  35.15  &  29.30  &  27.70  & --- & --- & --- & --- & --- & --- & --- & --- & ---\\ 
+0.6,+0.7  &  55.41  &  43.74  &  39.95  &  31.11  &  28.42  &  22.20  &  18.39  &  13.19  & --- & --- & --- & --- & --- & --- & --- & --- & --- & ---\\ 
+0.5,+0.6  &  49.41  &  36.85  &  27.84  &  23.68  &  21.21  &  17.60  & --- & --- & --- & --- & --- & --- & --- & --- & --- & --- & --- & ---\\ 
+0.4,+0.5  &  39.94  &  29.50  &  22.39  &  21.34  &  15.48  & --- & --- & --- & --- & --- & --- & --- & --- & --- & --- & --- & --- & ---\\ 
+0.3,+0.4  &  33.07  &  22.91  &  17.52  &  13.90  &  7.386  & --- & --- & --- & --- & --- & --- & --- & --- & --- & --- & --- & --- & ---\\ 
+0.2,+0.3  &  31.69  &  17.30  &  14.51  &  7.762  & --- & --- & --- & --- & --- & --- & --- & --- & --- & --- & --- & --- & --- & ---\\ 
+0.1,+0.2  &  27.32  &  12.88  &  9.982  & --- & --- & --- & --- & --- & --- & --- & --- & --- & --- & --- & --- & --- & --- & ---\\ 
0.0,+0.1   &  23.21  &  12.79  &  7.316  & --- & --- & --- & --- & --- & --- & --- & --- & --- & --- & --- & --- & --- & --- & ---\\ 
-0.1,0.0   &  21.26  &  8.773  & --- & --- & --- & --- & --- & --- & --- & --- & --- & --- & --- & --- & --- & --- & --- & ---\\ 
-0.2,-0.1  &  17.65  &  8.052  & --- & --- & --- & --- & --- & --- & --- & --- & --- & --- & --- & --- & --- & --- & --- & ---\\ 
-0.3,-0.2  &  18.40  & --- & --- & --- & --- & --- & --- & --- & --- & --- & --- & --- & --- & --- & --- & --- & --- & ---\\ 
-0.4,-0.3  &  17.97  & --- & --- & --- & --- & --- & --- & --- & --- & --- & --- & --- & --- & --- & --- & --- & --- & ---\\ 
-0.5,-0.4  &  14.18  & --- & --- & --- & --- & --- & --- & --- & --- & --- & --- & --- & --- & --- & --- & --- & --- & ---\\ 
-0.6,-0.5  &  16.58  & --- & --- & --- & --- & --- & --- & --- & --- & --- & --- & --- & --- & --- & --- & --- & --- & ---\\ 
-0.7,-0.6  & --- & --- & --- & --- & --- & --- & --- & --- & --- & --- & --- & --- & --- & --- & --- & --- & --- & ---\\ 
-0.8,-0.7  & --- & --- & --- & --- & --- & --- & --- & --- & --- & --- & --- & --- & --- & --- & --- & --- & --- & ---\\ 
-0.9,-0.8  & --- & --- & --- & --- & --- & --- & --- & --- & --- & --- & --- & --- & --- & --- & --- & --- & --- & ---\\ 
-1.0,-0.9  & --- & --- & --- & --- & --- & --- & --- & --- & --- & --- & --- & --- & --- & --- & --- & --- & --- & ---\\ 
\hline
\hline
\end{tabular}
\end{center}
\end{table*}

\begin{table*}
\begin{center}
\caption{
CCQE-like background in units of 10$^{-41}$~cm$^2$/GeV to the MiniBooNE $\numub$ CCQE double-differential cross section on mineral oil.  In this configuration, the hydrogen scattering component is treated as signal and is not included in the CCQE-like background.}
\label{tbl:dblDifflQElikeCH2}
\scriptsize\addtolength{\tabcolsep}{-0.8pt}
\begin{tabular}{c|cccccccccccccccccc}
\hline
\hline
\raisebox{-0.5ex}{$\cos\th_\mu$}\raisebox{0.5ex}{$T_\mu$(GeV)}&0.2,0.3&0.3,0.4&0.4,0.5&0.5,0.6&0.6,0.7&0.7,0.8&0.8,0.9&0.9,1.0&
                                              1.0,1.1&1.1,1.2&1.2,1.3&1.3,1.4&1.4,1.5&1.5,1.6&1.6,1.7&1.7,1.8&1.8,1.9&1.9,2.0\\
\hline
+0.9,+1.0  &  114.2  &  235.2  &  315.1  &  362.5  &  397.5  &  410.2  &  405.5  &  375.6  &  333.6  &  283.6  &  226.5  &  176.2  &  130.3  &  90.65 &---&---&---&---\\ 
+0.8,+0.9  &  91.75  &  170.2  &  194.0  &  190.6  &  177.6  &  148.7  &  119.5  &  91.94  &  61.22  &  40.46  &---&---&---&---&---&---&---&---\\ 
+0.7,+0.8  &  67.57  &  110.1  &  110.0  &  91.36  &  67.25  &  44.58  &  28.53  &---&---&---&---&---&---&---&---&---&---&---\\ 
+0.6,+0.7  &  48.98  &  70.90  &  60.55  &  40.69  &  24.32  &---&---&---&---&---&---&---&---&---&---&---&---&---\\ 
+0.5,+0.6  &  35.74  &  43.29  &  31.14  &  17.10  &---&---&---&---&---&---&---&---&---&---&---&---&---&---\\ 
+0.4,+0.5  &  25.59  &  27.80  &  15.67  &---&---&---&---&---&---&---&---&---&---&---&---&---&---&---\\ 
+0.3,+0.4  &  18.96  &  17.05  &---&---&---&---&---&---&---&---&---&---&---&---&---&---&---&---\\ 
+0.2,+0.3  &  12.54  &  9.613  &---&---&---&---&---&---&---&---&---&---&---&---&---&---&---&---\\ 
+0.1,+0.2  &---&---&---&---&---&---&---&---&---&---&---&---&---&---&---&---&---&---\\ 
0.0,+0.1  &---&---&---&---&---&---&---&---&---&---&---&---&---&---&---&---&---&---\\ 
-0.1,0.0  &---&---&---&---&---&---&---&---&---&---&---&---&---&---&---&---&---&---\\ 
-0.2,-0.1  &---&---&---&---&---&---&---&---&---&---&---&---&---&---&---&---&---&---\\ 
-0.3,-0.2  &---&---&---&---&---&---&---&---&---&---&---&---&---&---&---&---&---&---\\ 
-0.4,-0.3  &---&---&---&---&---&---&---&---&---&---&---&---&---&---&---&---&---&---\\
-0.5,-0.4  &---&---&---&---&---&---&---&---&---&---&---&---&---&---&---&---&---&---\\ 
-0.6,-0.5  &---&---&---&---&---&---&---&---&---&---&---&---&---&---&---&---&---&---\\ 
-0.7,-0.6  &---&---&---&---&---&---&---&---&---&---&---&---&---&---&---&---&---&---\\ 
-0.8,-0.7  &---&---&---&---&---&---&---&---&---&---&---&---&---&---&---&---&---&---\\ 
-0.9,-0.8  &---&---&---&---&---&---&---&---&---&---&---&---&---&---&---&---&---&---\\ 
-1.0,-0.9  &---&---&---&---&---&---&---&---&---&---&---&---&---&---&---&---&---&---\\ 
\hline
\hline
\end{tabular}
\end{center}
\end{table*}

\begin{table*}
\begin{center}
\caption{
The predicted CC1$\pim$ background in units of 10$^{-41}$~cm$^2$/GeV.  As described in Section~\ref{sbsec:cons}, this background corresponds to an empirical adjustment of the Rein-Sehgal~\cite{R-S} calculation based on the MiniBooNE CC1$\pip$ data~\cite{qePRD}.}
\label{tbl:dblDifflCCpim}
\scriptsize\addtolength{\tabcolsep}{-0.8pt}
\begin{tabular}{c|cccccccccccccccccc}
\hline
\hline
\raisebox{-0.5ex}{$\cos\th_\mu$}\raisebox{0.5ex}{$T_\mu$(GeV)}&0.2,0.3&0.3,0.4&0.4,0.5&0.5,0.6&0.6,0.7&0.7,0.8&0.8,0.9&0.9,1.0&
                                              1.0,1.1&1.1,1.2&1.2,1.3&1.3,1.4&1.4,1.5&1.5,1.6&1.6,1.7&1.7,1.8&1.8,1.9&1.9,2.0\\
\hline
+0.9,+1  &  85.72  &  183.4  &  250.1  &  288.5  &  316.7  &  321.5  &  311.6  &  282.6  &  242.8  &  199.2  &  153.7  &  114.4  &  57.10  &  6.809 &---&---&---&---\\ 
+0.8,+0.9  &  69.90  &  133.9  &  153.9  &  150.5  &  138.5  &  113.4  &  88.08  &  64.12  &  39.46  &  11.47  &  ---  &  ---  &  ---  &  ---  & --- & --- & --- & ---\\ 
+0.7,+0.8  &  49.96  &  84.65  &  85.19  &  70.93  &  50.74  &  31.92  &  17.41  &  ---  &  ---  &  ---  & --- & --- & --- & --- & --- & --- & --- & ---\\ 
+0.6,+0.7  &  35.68  &  53.85  &  45.96  &  30.16  &  15.68  &  ---  &  ---  &  ---  & --- & --- & --- & --- & --- & --- & --- & --- & --- & ---\\ 
+0.5,+0.6  &  25.57  &  31.62  &  22.11  &  10.67  &  ---  &  ---  & --- & --- & --- & --- & --- & --- & --- & --- & --- & --- & --- & ---\\ 
+0.4,+0.5  &  16.70  &  19.44  &  9.089  &  ---  &  ---  & --- & --- & --- & --- & --- & --- & --- & --- & --- & --- & --- & --- & ---\\ 
+0.3,+0.4  &  3.479  &  9.247  &  ---  &  ---  &  ---  & --- & --- & --- & --- & --- & --- & --- & --- & --- & --- & --- & --- & ---\\ 
+0.2,+0.3  &  0.222  &  0.567  &  --- & --- & --- & --- & --- & --- & --- & --- & --- & --- & --- & --- & --- & --- & --- & ---\\ 
+0.1,+0.2  &  ---  &  ---  &  ---  & --- & --- & --- & --- & --- & --- & --- & --- & --- & --- & --- & --- & --- & --- & ---\\ 
0.0,+0.1  &  ---  &  ---  &  ---  & --- & --- & --- & --- & --- & --- & --- & --- & --- & --- & --- & --- & --- & --- & ---\\ 
-0.1,0.0  & --- & --- & --- & --- & --- & --- & --- & --- & --- & --- & --- & --- & --- & --- & --- & --- & --- & ---\\ 
-0.2,-0.1  &  ---  & --- & --- & --- & --- & --- & --- & --- & --- & --- & --- & --- & --- & --- & --- & --- & --- & ---\\ 
-0.3,-0.2  & --- & --- & --- & --- & --- & --- & --- & --- & --- & --- & --- & --- & --- & --- & --- & --- & --- & ---\\ 
-0.4,-0.3  & --- & --- & --- & --- & --- & --- & --- & --- & --- & --- & --- & --- & --- & --- & --- & --- & --- & ---\\ 
-0.5,-0.4  & --- & --- & --- & --- & --- & --- & --- & --- & --- & --- & --- & --- & --- & --- & --- & --- & --- & ---\\ 
-0.6,-0.5  & --- & --- & --- & --- & --- & --- & --- & --- & --- & --- & --- & --- & --- & --- & --- & --- & --- & ---\\ 
-0.7,-0.6  & --- & --- & --- & --- & --- & --- & --- & --- & --- & --- & --- & --- & --- & --- & --- & --- & --- & ---\\ 
-0.8,-0.7  & --- & --- & --- & --- & --- & --- & --- & --- & --- & --- & --- & --- & --- & --- & --- & --- & --- & ---\\ 
-0.9,-0.8  & --- & --- & --- & --- & --- & --- & --- & --- & --- & --- & --- & --- & --- & --- & --- & --- & --- & ---\\ 
-1.0,-0.9  & --- & --- & --- & --- & --- & --- & --- & --- & --- & --- & --- & --- & --- & --- & --- & --- & --- & ---\\ 
\hline
\hline
\end{tabular}
\end{center}
\end{table*}


The single-differential cross-section $\frac{d\sigma}{d\qsqqe}$ measurement with shape uncertainty and CCQE-like background is given in Table~\ref{tbl:qsqqeCH2}.

\begin{table*}
\begin{center}
\caption{
The MiniBooNE $\numub$ CCQE single differential cross section $\frac{d^2\sigma}{d\qsqqe}$ on mineral oil, shape error, and the predicted CCQE-like and CC1$\pim$ backgrounds in units of cm$^{2}$/GeV$^{2}$. The total normalization error is 13.0\%.}
\label{tbl:qsqqeCH2}
\begin{tabular}{ccccccccc}
\hline
\hline
\multirow{2}{*}{$\qsqqe$ (GeV$^{2}$)} && \multirow{2}{*}{$\frac{d\sigma}{d\qsqqe}$} & \multirow{2}{*}{shape uncertainty} & CCQE-like subtracted & CC1$\pim$ subtracted\\
&& &  & background & background\\
\hline
0.00-0.05	&&$8.262\times 10^{-39}$	&$7.156\times 10^{-40}$&$4.400\times 10^{-39}$&$3.425\times 10^{-39}$ \\ 
0.05-0.10	&&$9.075\times 10^{-39}$	&$3.976\times 10^{-40}$&$3.023\times 10^{-39}$&$2.353\times 10^{-39}$ \\ 
0.10-0.15	&&$7.343\times 10^{-39}$	&$1.921\times 10^{-40}$&$1.920\times 10^{-39}$&$1.473\times 10^{-39}$ \\ 
0.15-0.20	&&$5.867\times 10^{-39}$	&$1.498\times 10^{-40}$&$1.297\times 10^{-39}$&$9.852\times 10^{-40}$ \\ 
0.20-0.25	&&$4.569\times 10^{-39}$	&$1.633\times 10^{-40}$&$8.972\times 10^{-40}$&$6.747\times 10^{-40}$ \\ 
0.25-0.30	&&$3.400\times 10^{-39}$	&$1.613\times 10^{-40}$&$6.183\times 10^{-40}$&$4.559\times 10^{-40}$ \\ 
0.30-0.35	&&$2.610\times 10^{-39}$	&$1.403\times 10^{-40}$&$4.397\times 10^{-40}$&$3.169\times 10^{-40}$ \\ 
0.35-0.40	&&$2.083\times 10^{-39}$	&$1.592\times 10^{-40}$&$3.126\times 10^{-40}$&$2.202\times 10^{-40}$ \\ 
0.40-0.45	&&$1.617\times 10^{-39}$	&$1.706\times 10^{-40}$&$2.260\times 10^{-40}$&$1.563\times 10^{-40}$ \\ 
0.45-0.50	&&$1.276\times 10^{-39}$	&$1.447\times 10^{-40}$&$1.661\times 10^{-40}$&$1.120\times 10^{-40}$ \\ 
0.50-0.60	&&$8.978\times 10^{-40}$	&$1.204\times 10^{-40}$&$1.081\times 10^{-40}$&$6.994\times 10^{-41}$ \\ 
0.60-0.70	&&$5.394\times 10^{-40}$	&$1.042\times 10^{-40}$&$6.207\times 10^{-41}$&$3.758\times 10^{-41}$ \\ 
0.70-0.80	&&$3.416\times 10^{-40}$	&$8.790\times 10^{-41}$&$3.882\times 10^{-41}$&$2.095\times 10^{-41}$ \\ 
0.80-1.00	&&$1.901\times 10^{-40}$	&$6.319\times 10^{-41}$&$2.119\times 10^{-41}$&$7.650\times 10^{-42}$ \\ 
1.00-1.20	&&$8.276\times 10^{-41}$	&$4.100\times 10^{-41}$&$9.146\times 10^{-42}$&$7.006\times 10^{-43}$ \\ 
1.20-1.50	&&$2.870\times 10^{-41}$	&$2.086\times 10^{-41}$&$2.368\times 10^{-42}$&$1.682\times 10^{-44}$ \\ 
1.50-2.00	&&$7.225\times 10^{-42}$	&$9.554\times 10^{-42}$&$1.037\times 10^{-43}$&$1.401\times 10^{-45}$ \\ 
\hline
\hline
\end{tabular}
\end{center}
\end{table*}


The total cross section including the hydrogen content is given in Table~\ref{tbl:absXsecCH2}.  As discussed in Appendix~\ref{apndx:modDepCH2}, these results are dependent on both the $\enuqe$ reconstruction assumptions (Equation~\ref{eqn:EnuQE}) and the nuclear model used.  To recognize these dependencies, the neutrino energy is labeled here as ``$E_{\nu}^{\textrm{QE,RFG}}$".

\begin{table*}
\begin{center}
\caption{
The MiniBooNE $\numub$ CCQE total cross section on mineral oil, errors, and 
predicted CCQE-like and CC1$\pim$ backgrounds in bins of $E_\nu^{\textrm{QE,RFG}}$ and units of cm$^2$.}
\label{tbl:absXsecCH2}
\begin{tabular}{cccccccc}
\hline
\hline
$E_\nu^{\textrm{QE,RFG}}$~(GeV)&&\,\,\,\,\,\,\,$\si$\,\,\,\,\,\,\,&\,\,\,\,\,\,\,\,\,\,\,shape error\,\,\,\,\,\,\,\,\,\,\,&\,\,\,\,\,\,\,total error\,\,\,\,\,\,\,&CCQE-like background & CC1$\pim$ background \\
\hline
0.40-0.45&&	$1.738\times 10^{-39}$	&	$3.433\times 10^{-40}$	&	$3.433\times 10^{-40}$	&	$5.199\times 10^{-40}$	&	$3.904\times 10^{-40}$ \\
0.45-0.50&&	$1.881\times 10^{-39}$	&	$3.097\times 10^{-40}$	&	$3.097\times 10^{-40}$	&	$6.108\times 10^{-40}$	&	$4.671\times 10^{-40}$ \\
0.50-0.55&&	$2.078\times 10^{-39}$	&	$3.178\times 10^{-40}$	&	$3.178\times 10^{-40}$	&	$6.752\times 10^{-40}$	&	$5.219\times 10^{-40}$ \\
0.55-0.60&&	$2.308\times 10^{-39}$	&	$3.139\times 10^{-40}$	&	$3.213\times 10^{-40}$	&	$7.205\times 10^{-40}$	&	$5.597\times 10^{-40}$ \\
0.60-0.65&&	$2.542\times 10^{-39}$	&	$2.901\times 10^{-40}$	&	$3.155\times 10^{-40}$	&	$7.534\times 10^{-40}$	&	$5.862\times 10^{-40}$ \\
0.65-0.70&&	$2.753\times 10^{-39}$	&	$3.282\times 10^{-40}$	&	$3.411\times 10^{-40}$	&	$7.799\times 10^{-40}$	&	$6.068\times 10^{-40}$ \\
0.70-0.75&&	$2.932\times 10^{-39}$	&	$2.736\times 10^{-40}$	&	$3.247\times 10^{-40}$	&	$8.048\times 10^{-40}$	&	$6.263\times 10^{-40}$ \\
0.75-0.80&&	$3.098\times 10^{-39}$	&	$2.897\times 10^{-40}$	&	$3.419\times 10^{-40}$	&	$8.334\times 10^{-40}$	&	$6.479\times 10^{-40}$ \\
0.80-0.90&&	$3.374\times 10^{-39}$	&	$2.705\times 10^{-40}$	&	$3.604\times 10^{-40}$	&	$8.848\times 10^{-40}$	&	$6.834\times 10^{-40}$ \\
0.90-1.00&&	$3.780\times 10^{-39}$	&	$2.426\times 10^{-40}$	&	$4.167\times 10^{-40}$	&	$9.517\times 10^{-40}$	&	$7.222\times 10^{-40}$ \\
1.00-1.10&&	$4.171\times 10^{-39}$	&	$2.486\times 10^{-40}$	&	$4.972\times 10^{-40}$	&	$1.017\times 10^{-39}$	&	$7.573\times 10^{-40}$ \\
1.10-1.30&&	$4.631\times 10^{-39}$	&	$3.799\times 10^{-40}$	&	$7.340\times 10^{-40}$	&	$1.087\times 10^{-39}$	&	$7.808\times 10^{-40}$ \\
1.30-1.50&&	$5.510\times 10^{-39}$	&	$8.591\times 10^{-40}$	&	$1.297\times 10^{-40}$	&	$1.261\times 10^{-39}$	&	$8.469\times 10^{-40}$ \\
1.50-2.00&&	$6.654\times 10^{-39}$	&	$1.911\times 10^{-39}$	&	$2.407\times 10^{-40}$	&	$1.388\times 10^{-39}$	&	$8.461\times 10^{-40}$ \\
\hline
\hline
\end{tabular}
\end{center}
\end{table*}

\subsection{\label{apndx:sbsec:12c} Cross-section results on $^{12}$C}

The tables in this section report the cross-section results reliant on the RFG subtract hydrogen CCQE events from the data.  In addition to the shape errors provided with each measurement, a normalization uncertainty of 17.4\% is applicable here.

The MiniBooNE $\numub$ CCQE double-differential cross section, shape uncertainty, and CCQE-like cross section treating the hydrogen CCQE content as background are given in Tables~\ref{tbl:dblDiffl12C},~\ref{tbl:dblDifflShape12C}, and~\ref{tbl:dblDifflQElike12C}, respectively.  Tables~\ref{tbl:qsqqe12C} and~\ref{tbl:absXsec12C} provide the same information for the single-differential and total cross section.

\begin{table*}
\begin{center}
\caption{
The MiniBooNE $\numub$ CCQE double-differential cross section on carbon.  The units are $10^{-41}~\ucmt/\uGeV$.}
\label{tbl:dblDiffl12C}
\scriptsize\addtolength{\tabcolsep}{-0.8pt}
\begin{tabular}{c|cccccccccccccccccc}
\hline
\hline
\raisebox{-0.5ex}{$\uz$}\raisebox{0.5ex}{$T_\mu$(GeV)}&0.2,0.3&0.3,0.4&0.4,0.5&0.5,0.6&0.6,0.7&0.7,0.8&0.8,0.9&0.9,1.0&
                                              1.0,1.1&1.1,1.2&1.2,1.3&1.3,1.4&1.4,1.5&1.5,1.6&1.6,1.7&1.7,1.8&1.8,1.9&1.9,2.0\\
\hline

+0.9,+1.0  &  213.4  &  311.5  &  517.6  &  710.8  &  848.8  &  969.0  &  1015  &  998.8  &  973.0  &  875.0  &  726.3  &  638.5  &  547.9  &  484.0  & --- & --- & --- & ---\\ 
+0.8,+0.9  &  301.1  &  448.9  &  664.0  &  800.9  &  839.5  &  743.8  &  667.6  &  518.4  &  389.7  &  238.6  &  126.7  &  92.13  & --- & --- & --- & --- & --- & ---\\ 
+0.7,+0.8  &  304.7  &  406.6  &  535.9  &  521.4  &  456.9  &  355.0  &  233.2  &  135.1  &  61.71  & --- & --- & --- & --- & --- & --- & --- & --- & ---\\ 
+0.6,+0.7  &  295.3  &  343.1  &  362.1  &  322.4  &  228.8  &  129.0  &  59.01  & --- & --- & --- & --- & --- & --- & --- & --- & --- & --- & ---\\ 
+0.5,+0.6  &  294.6  &  299.5  &  238.3  &  178.4  &  109.0  &  47.10  & --- & --- & --- & --- & --- & --- & --- & --- & --- & --- & --- & --- \\ 
+0.4,+0.5  &  228.6  &  214.0  &  145.7  &  88.79  &  34.79  & --- & --- & --- & --- & --- & --- & --- & --- & --- & --- & --- & --- & --- \\ 
+0.3,+0.4  &  178.9  &  143.2  &  94.80  &  33.00  & --- & --- & --- & --- & --- & --- & --- & --- & --- & --- & --- & --- & --- & --- \\ 
+0.2,+0.3  &  156.9  &  83.82  &  34.40  &  11.86  & --- & --- & --- & --- & --- & --- & --- & --- & --- & --- & --- & --- & --- & --- \\ 
+0.1,+0.2  &  103.6  &  52.63  &  17.23  & --- & --- & --- & --- & --- & --- & --- & --- & --- & --- & --- & --- & --- & --- & --- \\ 
0.0,+0.1  &  89.47  &  47.39  &  8.768  & --- & --- & --- & --- & --- & --- & --- & --- & --- & --- & --- & --- & --- & --- & --- \\ 
-0.1,0.0  &  80.99  &  19.33  & --- & --- & --- & --- & --- & --- & --- & --- & --- & --- & --- & --- & --- & --- & --- & --- \\ 
-0.2,-0.1  &  53.95  &  12.83  & --- & --- & --- & --- & --- & --- & --- & --- & --- & --- & --- & --- & --- & --- & --- & --- \\ 
-0.3,-0.2  &  30.79  & --- & --- & --- & --- & --- & --- & --- & --- & --- & --- & --- & --- & --- & --- & --- & --- & --- \\ 
-0.4,-0.3  &  24.58  & --- & --- & --- & --- & --- & --- & --- & --- & --- & --- & --- & --- & --- & --- & --- & --- & --- \\ 
-0.5,-0.4  &  27.97  & --- & --- & --- & --- & --- & --- & --- & --- & --- & --- & --- & --- & --- & --- & --- & --- & --- \\ 
-0.6,-0.5  &  24.58  & --- & --- & --- & --- & --- & --- & --- & --- & --- & --- & --- & --- & --- & --- & --- & --- & --- \\ 
-0.7,-0.6  & --- & --- & --- & --- & --- & --- & --- & --- & --- & --- & --- & --- & --- & --- & --- & --- & --- & --- \\ 
-0.8,-0.7  & --- & --- & --- & --- & --- & --- & --- & --- & --- & --- & --- & --- & --- & --- & --- & --- & --- & --- \\ 
-0.9,-0.8  & --- & --- & --- & --- & --- & --- & --- & --- & --- & --- & --- & --- & --- & --- & --- & --- & --- & --- \\ 
-1.0,-0.9  & --- & --- & --- & --- & --- & --- & --- & --- & --- & --- & --- & --- & --- & --- & --- & --- & --- & --- \\ 
\hline
\hline
\end{tabular}
\end{center}
\end{table*}

\begin{table*}
\begin{center}
\caption{
Shape uncertainty for the $\numub$ CCQE double-differential cross section on carbon.  The units are $10^{-41}~\ucmt/\uGeV$, and the total normalization uncertainty is 17.4\%.}
\label{tbl:dblDifflShape12C}
\scriptsize\addtolength{\tabcolsep}{-0.8pt}
\begin{tabular}{c|cccccccccccccccccc}
\hline
\hline
\raisebox{-0.5ex}{$\uz$}\raisebox{0.5ex}{$T_\mu$(GeV)}&0.2,0.3&0.3,0.4&0.4,0.5&0.5,0.6&0.6,0.7&0.7,0.8&0.8,0.9&0.9,1.0&
                                              1.0,1.1&1.1,1.2&1.2,1.3&1.3,1.4&1.4,1.5&1.5,1.6&1.6,1.7&1.7,1.8&1.8,1.9&1.9,2.0\\
\hline

+0.9,+1.0  &  98.14  &  100.4  &  123.6  &  137.5  &  133.7  &  125.3  &  120.8  &  118.1  &  134.4  &  146.6  &  152.9  &  173.4  &  193.8  &  309.0  &---&---&---&---\\ 
+0.8,+0.9  &  90.80  &  98.31  &  105.9  &  103.8  &  111.1  &  79.84  &  72.10  &  70.57  &  76.04  &  82.74  &  57.54  &  59.06  & --- & --- & --- & --- & --- & ---\\ 
+0.7,+0.8  &  79.23  &  75.16  &  78.05  &  69.10  &  59.27  &  50.25  &  46.95  &  40.67  &  34.90  & --- & --- & --- & --- & --- & --- & --- & --- & ---\\ 
+0.6,+0.7  &  73.98  &  59.60  &  53.93  &  45.68  &  39.91  &  31.08  &  36.83  & --- & --- & --- & --- & --- & --- & --- & --- & --- & --- & ---\\ 
+0.5,+0.6  &  68.65  &  50.39  &  37.82  &  32.77  &  28.89  &  22.26  & --- & --- & --- & --- & --- & --- & --- & --- & --- & --- & --- & ---\\ 
+0.4,+0.5  &  53.34  &  39.54  &  29.94  &  30.13  &  17.86  & --- & --- & --- & --- & --- & --- & --- & --- & --- & --- & --- & --- & ---\\ 
+0.3,+0.4  &  46.12  &  30.88  &  24.09  &  28.23  & --- & --- & --- & --- & --- & --- & --- & --- & --- & --- & --- & --- & --- & ---\\ 
+0.2,+0.3  &  42.37  &  23.09  &  19.04  &  11.05  & --- & --- & --- & --- & --- & --- & --- & --- & --- & --- & --- & --- & --- & ---\\ 
+0.1,+0.2  &  36.75  &  17.43  &  13.78  & --- & --- & --- & --- & --- & --- & --- & --- & --- & --- & --- & --- & --- & --- & ---\\ 
0.0,+0.1   &  31.26  &  18.63  &  10.54  & --- & --- & --- & --- & --- & --- & --- & --- & --- & --- & --- & --- & --- & --- & ---\\ 
-0.1,0.0   &  28.79  &  12.37  & --- & --- & --- & --- & --- & --- & --- & --- & --- & --- & --- & --- & --- & --- & --- & ---\\ 
-0.2,-0.1  &  23.23  &  10.23  & --- & --- & --- & --- & --- & --- & --- & --- & --- & --- & --- & --- & --- & --- & --- & ---\\ 
-0.3,-0.2  &  18.43  & --- & --- & --- & --- & --- & --- & --- & --- & --- & --- & --- & --- & --- & --- & --- & --- & ---\\ 
-0.4,-0.3  &  23.99  & --- & --- & --- & --- & --- & --- & --- & --- & --- & --- & --- & --- & --- & --- & --- & --- & ---\\ 
-0.5,-0.4  &  20.26  & --- & --- & --- & --- & --- & --- & --- & --- & --- & --- & --- & --- & --- & --- & --- & --- & ---\\ 
-0.6,-0.5  &  25.09  & --- & --- & --- & --- & --- & --- & --- & --- & --- & --- & --- & --- & --- & --- & --- & --- & ---\\ 
-0.7,-0.6  & --- & --- & --- & --- & --- & --- & --- & --- & --- & --- & --- & --- & --- & --- & --- & --- & --- & --- \\ 
-0.8,-0.7  & --- & --- & --- & --- & --- & --- & --- & --- & --- & --- & --- & --- & --- & --- & --- & --- & --- & --- \\ 
-0.9,-0.8  & --- & --- & --- & --- & --- & --- & --- & --- & --- & --- & --- & --- & --- & --- & --- & --- & --- & --- \\ 
-1.0,-0.9  & --- & --- & --- & --- & --- & --- & --- & --- & --- & --- & --- & --- & --- & --- & --- & --- & --- & --- \\ 
\hline
\hline
\end{tabular}
\end{center}
\end{table*}

\begin{table*}
\begin{center}
\caption{
The QE-like $\numub$ background subtracted from the double-differential cross section on carbon.  The $\numub$ CCQE interactions with hydrogen are treated as background in this calculation, and so their contribution is included here. The units are $10^{-41}~\ucmt/\uGeV$.}
\label{tbl:dblDifflQElike12C}
\scriptsize\addtolength{\tabcolsep}{-0.8pt}
\begin{tabular}{c|cccccccccccccccccc}
\hline
\hline
\raisebox{-0.5ex}{$\uz$}\raisebox{0.5ex}{$T_\mu$(GeV)}&0.2,0.3&0.3,0.4&0.4,0.5&0.5,0.6&0.6,0.7&0.7,0.8&0.8,0.9&0.9,1.0&
                                              1.0,1.1&1.1,1.2&1.2,1.3&1.3,1.4&1.4,1.5&1.5,1.6&1.6,1.7&1.7,1.8&1.8,1.9&1.9,2.0\\
\hline
+0.9,+1.0  &  222.1  &  409.8  &  555.9  &  660.5  &  736.9  &  766.0  &  752.7  &  689.3  &  601.6  &  498.5  &  392.2  &  307.1  &  227.2  &  157.1  &--- &--- &--- &---\\ 
+0.8,+0.9  &  187.7  &  314.9  &  368.4  &  380.1  &  357.0  &  299.0  &  240.4  &  182.5  &  118.4  &  74.33  &  41.40  &--- &--- &--- &--- &--- &--- &---\\ 
+0.7,+0.8  &  142.5  &  211.1  &  217.0  &  190.8  &  151.8  &  103.8  &  65.15  &  37.80  &--- &--- &--- &--- &--- &--- &--- &--- &--- &---\\ 
+0.6,+0.7  &  107.9  &  144.5  &  128.6  &  94.78  &  62.25  &  35.90  &  17.88  &--- &--- &--- &--- &--- &--- &--- &--- &--- &--- &---\\ 
+0.5,+0.6  &  82.68  &  95.12  &  74.83  &  46.40  &  23.98  &--- &--- &--- &--- &--- &--- &--- &--- &--- &--- &--- &--- &---\\ 
+0.4,+0.5  &  62.92  &  62.88  &  41.11  &  22.22  &--- &--- &--- &--- &--- &--- &--- &--- &--- &--- &--- &--- &--- &---\\ 
+0.3,+0.4  &  47.91  &  41.44  &  23.83  &--- &--- &--- &--- &--- &--- &--- &--- &--- &--- &--- &--- &--- &--- &---\\ 
+0.2,+0.3  &  33.44  &  25.40  &  11.47  &--- &--- &--- &--- &--- &--- &--- &--- &--- &--- &--- &--- &--- &--- &---\\ 
+0.1,+0.2  &  23.84  &  16.09  &--- &--- &--- &--- &--- &--- &--- &--- &--- &--- &--- &--- &--- &--- &--- &---\\ 
0.0,+0.1  &  18.46  &  10.31  &--- &--- &--- &--- &--- &--- &--- &--- &--- &--- &--- &--- &--- &--- &--- &---\\ 
-0.1,0.0  &  13.22  &--- &--- &--- &--- &--- &--- &--- &--- &--- &--- &--- &--- &--- &--- &--- &--- &---\\ 
-0.2,-0.1  &--- &--- &--- &--- &--- &--- &--- &--- &--- &--- &--- &--- &--- &--- &--- &--- &--- &---\\ 
-0.3,-0.2  &--- &--- &--- &--- &--- &--- &--- &--- &--- &--- &--- &--- &--- &--- &--- &--- &--- &---\\ 
-0.4,-0.3  &--- &--- &--- &--- &--- &--- &--- &--- &--- &--- &--- &--- &--- &--- &--- &--- &--- &---\\ 
-0.5,-0.4  &--- &--- &--- &--- &--- &--- &--- &--- &--- &--- &--- &--- &--- &--- &--- &--- &--- &---\\ 
-0.6,-0.5  &--- &--- &--- &--- &--- &--- &--- &--- &--- &--- &--- &--- &--- &--- &--- &--- &--- &---\\ 
-0.7,-0.6  &--- &--- &--- &--- &--- &--- &--- &--- &--- &--- &--- &--- &--- &--- &--- &--- &--- &---\\ 
-0.8,-0.7  &--- &--- &--- &--- &--- &--- &--- &--- &--- &--- &--- &--- &--- &--- &--- &--- &--- &---\\ 
-0.9,-0.8  &--- &--- &--- &--- &--- &--- &--- &--- &--- &--- &--- &--- &--- &--- &--- &--- &--- &---\\ 
-1.0,-0.9  &--- &--- &--- &--- &--- &--- &--- &--- &--- &--- &--- &--- &--- &--- &--- &--- &--- &---\\ 
\hline
\hline
\end{tabular}
\end{center}
\end{table*}

\begin{table*}
\begin{center}
\caption{
The MiniBooNE $\numub$ CCQE single differential cross section $\frac{d^2\sigma}{d\qsqqe}$ on carbon, shape error, and CCQE-like background in units of cm$^{2}$/GeV$^{2}$.  The $\numub$ CCQE content is treated as background, and the total normalization error is 17.4\%.}
\label{tbl:qsqqe12C}
\begin{tabular}{cccccccc}
\hline
\hline
\multirow{2}{*}{$\qsqqe$ (GeV$^{2}$)} && \multirow{2}{*}{$\frac{d\sigma}{d\qsqqe}$} & \multirow{2}{*}{shape uncertainty} & \multirow{2}{*}{CCQE-like background}\\
&& & & \\
\hline
0.00-0.05&&$6.076\times 10^{-39}$&$1.002\times 10^{-39}$&$7.809\times 10^{-39}$ \\ 
0.05-0.10&&$8.769\times 10^{-39}$&$5.319\times 10^{-40}$&$5.774\times 10^{-39}$ \\ 
0.10-0.15&&$7.495\times 10^{-39}$&$2.847\times 10^{-40}$&$3.727\times 10^{-39}$ \\ 
0.15-0.20&&$6.202\times 10^{-39}$&$2.384\times 10^{-40}$&$2.569\times 10^{-39}$ \\ 
0.20-0.25&&$4.921\times 10^{-39}$&$2.580\times 10^{-40}$&$1.812\times 10^{-39}$ \\ 
0.25-0.30&&$3.691\times 10^{-39}$&$2.161\times 10^{-40}$&$1.284\times 10^{-39}$ \\ 
0.30-0.35&&$2.838\times 10^{-39}$&$1.918\times 10^{-40}$&$9.322\times 10^{-40}$ \\ 
0.35-0.40&&$2.297\times 10^{-39}$&$2.084\times 10^{-40}$&$6.851\times 10^{-40}$ \\ 
0.40-0.45&&$1.786\times 10^{-39}$&$2.270\times 10^{-40}$&$5.077\times 10^{-40}$ \\ 
0.45-0.50&&$1.418\times 10^{-39}$&$1.915\times 10^{-40}$&$3.841\times 10^{-40}$ \\ 
0.50-0.60&&$9.995\times 10^{-40}$&$1.538\times 10^{-40}$&$2.603\times 10^{-40}$ \\ 
0.60-0.70&&$5.981\times 10^{-40}$&$1.425\times 10^{-40}$&$1.557\times 10^{-40}$ \\ 
0.70-0.80&&$3.757\times 10^{-40}$&$1.220\times 10^{-40}$&$9.733\times 10^{-41}$ \\ 
0.80-1.00&&$2.104\times 10^{-40}$&$8.748\times 10^{-41}$&$5.225\times 10^{-41}$ \\ 
1.00-1.20&&$9.186\times 10^{-41}$&$7.001\times 10^{-41}$&$2.382\times 10^{-41}$ \\ 
1.20-1.50&&$3.099\times 10^{-41}$&$2.665\times 10^{-41}$&$1.014\times 10^{-41}$ \\ 
1.50-2.00&&$7.414\times 10^{-42}$&$1.267\times 10^{-41}$&$3.167\times 10^{-42}$ \\ 
\hline
\hline
\end{tabular}
\end{center}
\end{table*}

\begin{table*}
\begin{center}
\caption{
The MiniBooNE $\numub$ CCQE total cross section on carbon, errors, and 
predicted CCQE-like background in bins of $E_\nu^{\textrm{QE,RFG}}$ and units of cm$^2$.}
\label{tbl:absXsec12C}
\begin{tabular}{ccccccc}
\hline
\hline
$E_\nu^{\textrm{QE,RFG}}$~(GeV)&&$\si$&\,\,\,\,\,\,\, shape error\,\,\,\,\,\,\,&\,\,\,\,\,\,\,total error\,\,\,\,\,\,\,&CCQE-like background\\
\hline
0.40-0.45&&	$1.808\times 10^{-39}$	&	$6.267\times 10^{-40}$	&	$6.267\times 10^{-40}$	&	$1.127\times 10^{-39}$ \\
0.45-0.50&&	$1.890\times 10^{-39}$	&	$4.471\times 10^{-40}$	&	$4.471\times 10^{-40}$	&	$1.224\times 10^{-39}$ \\
0.50-0.55&&	$2.019\times 10^{-39}$	&	$4.359\times 10^{-40}$	&	$4.433\times 10^{-40}$	&	$1.309\times 10^{-39}$ \\
0.55-0.60&&	$2.258\times 10^{-39}$	&	$4.102\times 10^{-40}$	&	$4.384\times 10^{-40}$	&	$1.386\times 10^{-39}$ \\
0.60-0.65&&	$2.501\times 10^{-39}$	&	$3.761\times 10^{-40}$	&	$4.335\times 10^{-40}$	&	$1.454\times 10^{-39}$ \\
0.65-0.70&&	$2.728\times 10^{-39}$	&	$4.209\times 10^{-40}$	&	$4.559\times 10^{-40}$	&	$1.512\times 10^{-39}$ \\
0.70-0.75&&	$2.932\times 10^{-39}$	&	$3.528\times 10^{-40}$	&	$4.390\times 10^{-40}$	&	$1.575\times 10^{-39}$ \\
0.75-0.80&&	$3.091\times 10^{-39}$	&	$3.574\times 10^{-40}$	&	$4.560\times 10^{-40}$	&	$1.645\times 10^{-39}$ \\
0.80-0.90&&	$3.372\times 10^{-39}$	&	$3.385\times 10^{-40}$	&	$4.821\times 10^{-40}$	&	$1.753\times 10^{-39}$ \\
0.90-1.00&&	$3.815\times 10^{-39}$	&	$3.195\times 10^{-40}$	&	$5.663\times 10^{-40}$	&	$1.895\times 10^{-39}$ \\
1.00-1.10&&	$4.254\times 10^{-39}$	&	$3.331\times 10^{-40}$	&	$6.704\times 10^{-40}$	&	$2.022\times 10^{-39}$ \\
1.10-1.30&&	$4.789\times 10^{-39}$	&	$5.207\times 10^{-40}$	&	$9.831\times 10^{-40}$	&	$2.121\times 10^{-39}$ \\
1.30-1.50&&	$5.784\times 10^{-39}$	&	$1.162\times 10^{-39}$	&	$1.742\times 10^{-39}$	&	$2.378\times 10^{-39}$ \\
1.50-2.00&&	$7.086\times 10^{-39}$	&	$2.440\times 10^{-39}$	&	$3.126\times 10^{-39}$	&	$2.482\times 10^{-39}$ \\
\hline
\hline
\end{tabular}
\end{center}
\end{table*}

\end{document}